\documentclass[numberedappendix]{emulateapj}
\slugcomment{ApJ, in press}

\shortauthors{Hwang \& Laming}

\begin{document}
\title{A Chandra X-ray Survey of Ejecta in the Cassiopeia A Supernova
  Remnant}


\author{Una Hwang\altaffilmark{1} \& J. Martin Laming\altaffilmark{2}}


\altaffiltext{1}{Goddard Space Flight Center and Johns Hopkins University\\
\email{Una.Hwang-1@gsfc.nasa.gov}}
\altaffiltext{2}{Code 7674L, Naval Research Laboratory, Washington DC 20375\\
\email{laming@nrl.navy.mil}}

\begin{abstract}
We present a survey of the X-ray emitting ejecta in the Cassiopeia A
supernova remnant based on an extensive analysis of over 6000 spectral
regions extracted on 2.5-10$''$ angular scales using the Chandra 1 Ms
observation.  We interpret these results in the context of
hydrodynamical models for the evolution of the remnant.  The
distributions of fitted temperature and ionization age, and the
implied mass coordinates, are highly peaked and suggest that the
ejecta were subjected to multiple secondary shocks following reverse
shock interaction with ejecta inhomogeneities.  Based on the fitted
emission measure and element abundances, and an estimate of the
emitting volume, we derive masses for the X-ray emitting ejecta and
also show the distribution of the mass of various elements over the
remnant. An upper limit to the total shocked Fe mass visible in X-rays
appears to be roughly 0.13 M$\odot$, which accounts for nearly all of
the mass expected in Fe ejecta.  We find two populations of Fe ejecta,
that associated with normal Si-burning and that possibly associated
with $\alpha$-rich freeze-out, with a mass ratio of approximately 2:1.
Essentially all of the observed Fe (both components) lies well outside
the central regions of the SNR, possibly having been ejected by
hydrodynamic instabilities during the explosion.  We discuss this, and
its implications for the neutron star kick.
\end{abstract}

\keywords{hydrodynamics -- ISM: individual (Cassiopeia A) -- supernova remnants -- X-rays: ISM}

\section{Introduction}

Theory and observations are merging toward consensus that
core-collapse supernova explosions are intrinsically asymmetric, even
if the progenitor had an initially symmetric configuration.  Recent
three-dimensional simulations show that strong dynamical interactions
between the various burning shells can lead to large asymmetries in
the progenitors, independent of the symmetry of the explosion
\citep{arnett10}.  Nor is the explosion itself likely to be spherical,
with convective instabilities \citep[e.g.][]{herant92, herant95} and
instabilities at the accretion shock \citep[e.g.][]{blondin03,
  foglizzo07} all possibly acting to create asymmetries.  It is
further anticipated that such asymmetries may hold the key to the
core-collapse explosion mechanism \citep[e.g.][]{ott08,brandt11}.
They will be manifested not only in the distributions of the ejecta
mass and velocity, but also in the recoil of the nascent neutron star,
which is observed with measured space velocities of up to 1000-1500
km/s \citep{arzoumanian02,hobbs05,faucher06}.  Simulations show that a
neutron star kick that has a purely hydrodynamic origin can balance
the total momentum of the anisotropic component, and is expected
in the direction opposite the fastest shock expansion
\citep{wongwat10, scheck04}.

Observations of core-collapse supernovae indicate a significant degree
of explosion asymmetry via measurements such as time-dependent
polarization changes or Doppler velocities \citep{mazzali06,
  leonard06}, but explosion asymmetries are also revealed in
sufficiently young supernova remnants (SNRs) through their
shock-heated ejecta.  The most detailed view so far available of the
ejecta in a young core-collapse SNR is provided by Cassiopeia A, which
at 330 years is the penultimate of the known Galactic supernova
remnants.  Its reverse shock has already progressed deeply into the
explosively produced nucleosynthesis products, aided by the strong
presupernova mass loss incurred by its progenitor through the likely action of a
binary companion \citep{young06}.

That Cas A was produced by an asymmetric explosion is by now
well-established, and most recently by light echo observations which
sample the explosion hundreds of years after the fact in different
directions.  These show variations in the ejecta velocities of
$\sim4000$ km/s (Rest et al. 2010).  Chandra X-ray observations have
also played an important role in revealing the complexity of Cas A's
X-ray emitting ejecta.  The X-ray emitting Si ejecta show a bipolar
structure with jet-like features \citep{hwang04, vink04, laming06}
similar to that seen in optical \citep[][and references
  therein]{fesen01} and infrared emission \citep{hines04}.  The X-ray
ejecta spectra reveal the imprint of roughly a factor of two
asymmetries in the deposition of explosion energy around the remnant
\citep{laming03}, similar to the distribution of kinetic energies and
positions of fast moving optical knots \citep{hammell08}. Early
Chandra observations identified regions dominated by emission from Fe
ejecta well outside the projected location of Si emission that were
proposed to be sites of overturn between the nucleosynthesis layers
\citep{hughes00}.  This conclusion is apparently supported by the
X-ray Doppler measurements showing higher velocities of Fe compared to
Si \citep{willingale02}.  More recently, the dynamical structure of
the ejecta has been inferred in remarkable detail by the
multi-wavelength (X-ray, optical, infrared) Doppler-shift maps
compiled by \citet{delaney10}.  These show the inner ejecta to be
unshocked and in a flattened distribution ([Si II]), outlying matter
to be arranged in rings on a spherical surface ([Ar II], [Ne II]) ,
and outflows emerging out of the surface of the remnant (Fe ejecta,
jet, and counterjet).  Their detailed analysis shows that the outflows
of X-ray emitting Fe ejecta are not material mixed outwards from
overturning of the ejecta layers, but rather, material that has pushed
through the overlying ejecta and is now encircled by rings of
material from outlying nucleosynthesis layers.

Large-scale studies of the X-ray emission in Cas A have already been
undertaken by \citet{stage06} and \citet{helder08}, but these
focus on the properties of the nonthermal emission.  The most
comprehensive accounting of the X-ray emitting ejecta in Cas A to date
was carried out by \citet{willingale03}, who used XMM-Newton CCD
observations to examine 225 regions of fixed 20$''$ size across Cas A
\citep{willingale02}.  They fitted a two-component thermal model
throughout to represent hot shocked circumstellar material plus cooler
shocked ejecta, and inferred a total ejecta mass of 2.2 M$_\odot$.
They interpret the Fe K emission as forming a bipolar double cone that
is associated with ejecta bullets that have broken beyond the forward
shock.

A detailed accounting of the X-ray emitting ejecta on fine angular
scales was one of the primary goals of a 1 Ms Chandra observation of
Cassiopeia A that was obtained in 2004, and is the subject of this
paper.  Our spectral survey includes over 6000 spectral regions
extracted on 2.5-10$''$ angular scales.  It is distinguished from
previous Chandra studies of the ejecta in Cas A, which have relied
either on spectral imaging or the detailed spectral analysis of a
relatively small number of regions.  We are able to cover a larger
area of the remnant with much higher sensitivity and angular
resolution than was available to \citet{willingale02}.  In particular,
we are able to show the distribution of the X-ray emitting Fe ejecta
in detail.  The Fe ejecta, being produced just outside the collapsing
core of the supernova, provide valuable clues to understand important
details of the explosion, including the recoil of the ejecta with the
neutron star.

We interpret the spectra in the context of the hydrodynamical models
applied by us previously to Cas A \citep{laming03, hwang03, laming06,
  hwang09}.  Cas A particularly lends itself to quantitative analysis
by virtue of its well-constrained age \citep{thorstensen01, fesen06}
and distance \citep{reed95}.  Recent one-dimensional hydrodynamical
models for the interaction of the remnant with a circumstellar wind
medium indicate that the reverse shock in Cas A has already interacted
with a significant fraction of the ejecta \citep{chevalier03}.
\citet{laming03} apply their models directly to Chandra X-ray spectra,
and also infer that there is very little of the unshocked ejecta
remaining.  Thus a large fraction of the ejecta in Cas A are
accessible to X-ray observations. The exceptions are the small amount
of ejecta that have not yet been shocked, ejecta at very low densities
(which we will discuss later), any ejecta that might have been subject
to rapid cooling by thermal instability, and and ejecta of very high
density that have been shocked to lower temperatures and thus emit at
longer wavelengths as optical knots \citep[e.g.][]{fesen01}.

The organization of the paper is as follows: Section 2 gives an
overview of the analysis procedure, with explanations as to the
spectral grid and background, and most importantly how the spectra
were organized according to the fitting model required.  We then
proceed to discuss the results, from maps for the remnant as a whole
to a few observations about properties of the forward shock-dominated
regions, and then of the ejecta.  Section 3 discusses issues concerning
the models and the refinements that have been made for this work
compared to previous work.  Section 4 then presents the results of the
mass calculation and issues connected to it, such as the presence of
unshocked ejecta and implications for the neutron star kick.

\section{Analysis Procedure}

We use the 1 Ms Advanced CCD Imaging Spectrometer (ACIS) observation
obtained by the Chandra X-ray Observatory in February and May 2004
(one of the first Chandra VLP, or Very Long Project, observations).
This observation is described by \citet{hwang04}.  Its most salient
features are that it was obtained in nine observation segments
(OBSIDS) and in GRADED mode, wherein the CCD detection events are
characterized onboard the spacecraft before telemetry.  It was
necessary to use GRADED mode to reduce the telemetry load, due to the
high source count rate (the observation accumulated some 280 million
photon events in 980 ks).  The main disadvantage of using GRADED mode
is that detailed corrections of detector problems such as pulse pile
up are not possible.  The data were processed with CALDB Version
3.2.2.

In this section, we outline the analysis procedure, which includes
definition of the spectral grid, considerations for the subtraction of
the background, the classification of forward shock and
ejecta-dominated spectra, the selection of appropriate spectral
models, and the results of the fitting.

\subsection{Spectral Grid}

To carry out our spectral survey of the ejecta in Cas A, we define a
grid of 10$''$ boxes across the entire supernova remnant (SNR),
optionally subdividing by factors of two down to 2.5$''$ depending on
the number of counts.  The regions are thus square or rectangular with
sides of length 2.5, 5, or 10$''$.  We excluded the regions dominated
by the central compact object, and
a few other regions with low numbers of counts.  This gave us 6202
spectra for analysis.  Although this is a very large number, the data
would have readily allowed at least a factor of two finer grid sizes
down to nearly the Chandra angular resolution in some regions.  This
being the first attempt at an analysis of the complex thermal emission
on such a fine angular scale, we tried to keep the number of grid
regions relatively tractable.

Each spectrum was then extracted individually from all nine OBSIDS
comprising the Ms observation, and the corresponding individual photon
redistribution matrices and effective area files computed.  The
individual spectra were then added, and the individual response files
weighted and added, to obtain the final spectral and response files
for each grid region.  The pulse-height spectra were further binned to
yield a minimum of 25 counts per bin to apply Gaussian statistics.

The filamentary and knotty characteristics of Cas A's X-ray emission,
its high surface brightness, and the excellence of the Chandra
mirrors, result in pulse pile-up at the brightest ejecta knots. Pulse
pile-up can produce spurious line features, for example, with energies
that correspond roughly to the summed energy of the Si and/or S He
$\alpha$ blends that are so prominent in Cas A.  Pile-up of continuum
photons will also skew the spectral shape and result in higher fitted
temperatures.

A detailed correction for pulse pile-up cannot be performed for these
GRADED mode data, and we made no attempt to account for these effects
for the thousands of spectra examined here.  We can, however, estimate
qualitatively the number of our spectra for which pile-up is a
significant problem.  For the {\it vpshock} fits described later (in section
2.3), we consider the distribution of the fitted temperature against
the counts per pixel for each spectrum.  The average fitted
temperature shows a clear trend of increasing for spectra with more
than about 6000 counts/pixel in 980 ks\footnote{The count/pixel
  cut-off we discuss here is of course specific for this particular
  observation.  For reference, the brightest pixels contain on the
  order of 16000 counts, corresponding to a count rate of 0.05 counts
  per 3.24 s frame. This corresponds roughly to pileup fractions of
  less than a few percent according to the Chandra Proposers'
  Observatory Guide.}.  For the 25 regions with the highest
count/pixel ratios ($>$ 6000 counts/pixel), the average fitted
temperature is 2.25 keV, compared to an average of 1.7 keV for the
remaining spectra.  Inspection of these spectra show the problems in
fitting the line emission described above; these problems are not
apparent for lower count-rate spectra.  Since the pile-up occurs
noticeably in only a limited number of regions ($\sim$25), we expect
that its uncorrected effect on the total mass is not large.

\subsection{Background}

The off-source regions outside the SNR show a line-rich spectrum
similar to that of the SNR itself.  The background is thus dominated
by the bright source spectrum due to scattering by intervening dust,
the CCD readout, and to a lesser extent, the wings of the mirror point
spread function.  Dust scattering has been shown to be significant for
Cas A based on ROSAT and Einstein observations by \citet{predehl95}
and \citet{mauche89}.

Since the source spectrum, and hence the background, varies both in
form and brightness around the remnant, we extract 16 separate
background spectra by subdividing a $0.4'$ thick, $\sim 3.5'$ radius
circular shell surrounding the remnant so that each segment covers
roughly 20 degrees in azimuth.  We then assign a background spectrum
for each source spectrum based on the source region's azimuthal
position angle.  Where the source extends beyond the background
extraction radius, near the northeast ``jet'' and its southwest
counterpart, we use the sum of the two nearest background regions to
either side.

We choose to subtract the background from the source spectrum rather
than to fit a model to each background spectrum.  A model for the
background would have to include a thermal component for the scattered
source spectrum, whose normalization would inevitably be difficult to
constrain.  The normalization for the particle background component is
well-constrained by the spectrum at high energies, but the effect of
scaling the scattered thermal background in the same way as the
particle background is similar to simply subtracting the total
background.

Simple subtraction of the off-source background means that the
background is generally underestimated, and by different amounts
depending on the photon energy and the dominant scattering process.
The precise determination of the background is thus a sophisticated
problem that is beyond the scope of our study.  The ejecta survey
which is the focus of our efforts will be less sensitive to the
precise background subtraction because most of the ejecta regions are
bright, with strong line emission; the subtleties of the background
subtraction are less important here than for inferences about the origin of
weak lines.



\subsection{Spectral Classification}

\subsubsection{Spectral Map}

To provide a consistent spectral characterization of the entire
remnant, we fit to every spectrum in our grid a single-component
plane-parallel shock model with variable element abundances (XSPEC
model {\em vpshock}; Borkowski et al. 2001) modified by interstellar
absorption (XSPEC model {\em wabs}; Morrison \& McCammon, 1983).  The
{\em vpshock} model incorporates ionization ages from zero up to a
fitted maximum value.  In previous work \citep{laming03,hwang03}, we
had used simple nonequilibrium ionization (NEI) models characterized
by a single temperature and ionization age, but there the extraction
regions were generally about 3$''$ and the exposure time only 50 ks.
We find that for the much longer exposure time of these data, and the
sometimes larger extraction regions, such models are no longer
acceptable.  Because most of the regions have ejecta-dominated
spectra, we have not included any elements lighter than O in the
model, taking O to be the primary source of the continuum in the
manner of \citet{vink96}, \citet{laming03} and \citet{hwang03}.  The
abundances of elements heavier than O are generally varied (with Ni
linked to Fe, Ca to Ar, and sometimes Ar to S), while O is held fixed
at the solar value, using the solar abundances of
\citet{anders89}\footnote{In practice, the emission is calculated as
  if all elements, specifically H, were included at their solar
  abundance ratios.  The individual contribution of each element is
  then further weighted by the abundance value relative to solar, so
  that those elements with zero abundance values do not contribute.}.
The redshift is allowed to vary to either positive or negative values,
given that the ejecta in Cas A have significant bulk motions
\citep[e.g.,][]{markert83, willingale02}, and is generally driven by
the Fe L emission and the strong Si blend.  Given that there may be
calibration problems near Si (see the discussion in DeLaney et
al. 2010), we consider the fitting of the redshift more as an aid to
achieving good fits rather than a reliable measure of the actual
line-of-sight velocity.  The fitted redshift does, however, reproduce
the gross features of the bulk motion as measured by other means.  A
Gaussian smoothing scale for the spectrum was also fitted, given that
the observed strong spectral lines are generally broad because of bulk
motions.

While we do not present errors for the fitted parameters here, these
were computed for key parameters.  We have found that this is a
necessary step to optimize finding the true minima: the $\chi^2$
terrain is generally rather rugged for these fits.  The redshift and
ionization age in particular are prone to settle at secondary minima,
as might be expected, considering that these parameters both have a
strong effect on the energies of the strong line features.

Here we note some of the general features of these results based on
images shown in Figure 1,
where fitted model parameters such as temperature, ionization age, and
various element abundances are shown for each spectral region.
Histograms and a scatter plot for the fitted temperatures and
ionization ages are shown in Figure 2.  The column density increases
systematically to the west by about a factor of 2 or more and reaches
a maximum at the western extremity of the remnant.  This is entirely
consistent with previous work \citep{keohane96, willingale02}.  There
are also smaller regions in the center with significant localized
column density enhancements of about 50\%.  The temperature
distribution has two peaks, which are seen most clearly in the
histogram in Figure 2.
The temperature map shows that the lower peak is associated with
ejecta-dominated, line-emitting regions and the higher with the forward
shock-dominated regions that were identified in previous work
\citep{gotthelf01, stage06, helder08}. (We will discuss the forward
shock in the following section.)  The ionization age, by contrast, is
remarkably narrowly distributed overall, though
coherent regions with high ionization ages can be identified.  In the
western half of the remnant, these high ionization regions appear to
be associated with the forward shock.  The highest ionization ages are
found in the east, however, and there they are clearly associated with
Fe-enriched ejecta, as is apparent from comparing the ionization age
and Fe abundance maps shown in Figure 1.

We will focus our attention primarily on the abundance maps for Fe and
Si, although such maps are also shown for all the elements for which
we fitted the abundances.  The maps for S and Ar are on the whole
similar to that of Si.  The abundance maps for Si and Fe are
distinctly different, but both feature three main lobes of ejecta
located north, east, and west.  The Si is also extended to the
northeast along the northeast ejecta ``jet'', while Fe is particularly
distinct in the eastern region compared to the Si, and shows its
highest enhancement in the outermost parts of this region.  In
general, the abundance maps we obtain are strongly reminiscent of the
corresponding line images shown by \citet{hwang00} and
\citet{hwang04}, and also resemble the less detailed abundance maps
obtained by \citet{willingale02}.  Those authors had previously noted
that the line emission and element abundance patterns for Si, S, Ar,
and Ca in Cas A are similar to each other.

The Ne and Mg maps have a distinct character in that they do not show
any prominent morphological characteristics aside from a brightening
at the western end of the remnant.  One must be cautious to interpret
those results, however, as the line emissions of Ne and Mg fall in
complicated parts of the spectrum, and the fitted abundances of these
elements may be correlated with parameters such as the ionization age,
which shows similar distribution patterns in the west, and the column
density, which is very high in that region.  It is clear, however,
that Ne and Mg both show a strikingly different morphology to Si, S,
and Ar, or to Fe, and are much more similar to each other than to any
of the other elements.


\subsubsection{Forward Shocked Regions}

Broadly speaking, the distinctions between ejecta- and FS-dominated
regions in Cas A are readily apparent, with differences in
temperature, ionization age, and element abundances.  To carry out a
survey of the ejecta mass, however, we must either model both the
ejecta and CSM components for each spectrum or else accurately
identify the specific regions where the reverse-shocked ejecta make
the dominant contribution to the emission.  Given the scope of the
spectral analysis, we have adopted the latter approach.
Multi-component fits can be difficult to constrain reliably,
particularly if one of the components is relatively weak, and thus
would require more individual attention than is feasible for a sample
of thousands.  Consequently, our next aim is to identify and eliminate
regions whose spectra can be completely associated with the forward
shock.

We evaluate the presence of thermal emission associated with the
forward shock by fitting a second set of plane-parallel shock models
to every spectrum, but this time with element abundances appropriate
for the CSM.  The optically emitting quasi-stationary flocculi (QSFs)
in Cas A are understood to be circumstellar mass loss from the
progenitor.  While abundance measurements for QSFs are limited to a
small number of knots, these generally show an order of magnitude
enhancement of N and sometimes also of He \citep{chevalier78}.
Theoretical calculations for the presupernova composition are also
given by \citet{arnett96}, where the models allow the elements H, He,
and N, all apparently present in Cas A, to exist simultaneously at a
narrow temperature range near log T (10$^9$ K) = -1.5 (their Figure
7.6).  At that temperature the abundance of He is 3 times the solar
value relative to H by number, and that of N about 15 times the solar
value.  As these abundances for He and N are broadly consistent with
the observational measurements, we proceed to adopt them for our fits,
along with solar values for the remaining elements, as representative
CSM element abundances.

About 1209 regions gave reasonably good fitting results ($\chi^2 \leq
1.2$) with the $vpshock$ model and these QSF element abundances, and
are thus assigned to the forward shock.  They are distributed mainly
in the remnant's outer rim and southwest interior, as would be
expected based on the 4-6 keV X-ray continuum image that highlights
the forward-shocked regions \citep{gotthelf01}.
Their average temperature is 2.2 keV, and their ionization ages are
rather narrowly distributed with an average value of $2\times 10^{11}$
cm$^{-3}$s.  These values can be assessed in the context of the models
of \citet{laming03}, which give the current density of the CSM at the
forward shock at about 1.5-2 cm$^{-3}$.  Considering the $r^{-2}$
dependence of the circumstellar density, the forward shock will have
encountered much denser material in the past and the present-day
ionization state of the forward shocked material is expected to be
relatively advanced.  The models give values of the ionization age in
the $10^{11}$ cm$^{-3}$ s range, approaching $10^{12}$ cm$^{-3}$s;
they also indicate that gas is rather hot, with temperatures from
2.5-4 keV.  This is reasonably close to the average values of the
temperature and ionization age that we find in our region, though the
fitted spectra don't show as broad a range in ionization age as is
predicted.

To the forward-shock regions identified solely by the thermal emission
model above, we must also add those that have a strong nonthermal
contribution.  To identify these, we devise a rough diagnostic for the
smoothness of the X-ray spectrum.  We bin each background-subtracted
spectrum at each significant line feature and continuum interval (some
of these cover only a narrow energy range), compute the ratio of
counts for each major line feature relative to counts in an adjacent
continuum bin, and take the sum of these ratios for all the line
features.  The distribution of this quantity for all the spectra has
two overlapping peaks; we take the spectra associated with the lower
peak (corresponding to weak lines in the spectrum) and perform further
fits with a composite plane-parallel pshock plus power-law model.  For
a cutoff in $\chi^2 \leq 1.2$ for these fits, we associate 206
additional regions with the forward shock, giving a total of 1415
regions with spectra that are consistent with emission that can be
associated with the forward shock alone.

Figure 3 summarizes the distribution and spectral characteristics of
the forward shock regions.  Their locations echo the 4-6 keV continuum
maps shown by \citet{gotthelf01} and \citet{hwang04}.  As already
noted, their temperatures and most especially their ionization ages
are rather narrowly distributed, more so than for the sample as a
whole.  From the Figure, the peaks in the distributions correspond
roughly to kT = 2.2 keV and log (n$_{\rm e}$ t) = 11.25, or n$_{\rm
  e}$t = 1.8e11 cm$^{-3}$s.  These regions will not be considered
further here as we focus our study on the ejecta hereafter.  We will
undertake a detailed consideration of the thermal emission associated
with the forward shock in subsequent work.

\subsection{Spectral Survey of the Ejecta and the Presence of Pure Fe}

We associate with ejecta the remaining, more than 4000, regions that
are inconsistent with forward shocked material.  For these regions,
the basic spectral model is the simple one-component plane-parallel
shock that has already been presented and discussed.  As noted above,
we take the view that, for the ejecta sample as a whole, it is
justifiable to assume sufficient ejecta dominance to neglect the
forward shock component.  Representative examples of all the various
types of spectra seen in Cas A are shown in Figure 5.

For some ejecta regions, the one-component spectral model is clearly
inadequate to describe the ejecta emission in that it fails to account
for the Fe K blend.  This has already been noted by \citet{hwang09}.
While there may be significant deficiencies in the atomic data that
are used for the thermal emission models, particularly in the
nonequilibrium ionization case, these cannot explain so large an
effect as we see.  In some cases, a strong Fe K blend is completely
unaccounted for by the model, which otherwise characterizes the
spectrum well.

Given the strong chemical inhomogeneities seen and expected in
core-collapse supernova remnants, it is plausible that Fe-rich ejecta
could be superposed along a given line of sight with ejecta of more
normal composition, and that each could have distinct plasma
conditions as well as abundances.  Moreover, nucleosynthesis models
predict the formation of nearly pure Fe ejecta by $\alpha$-rich
freeze-out during complete Si burning; this is in addition to ejecta
composed of a range of elements from Si and Fe formed by incomplete
Si-burning.

\citet{hwang03} and \citet{hwang04} have reported the presence of
highly enriched Fe ejecta in the southeastern region of Cas A, with
one region particularly appearing to be a``pure Fe'' cloud.  We take
another look at this region using the 1 Ms observation.  Figure 4
shows a portion of the southeastern region of Cas A and the roughly
2.5$''$ extraction region that we use for this cloud; this is smaller
than the region used by Hwang et al. 2003.  The background is taken
locally from the vicinity of the source region, and we compared
results using three different background spectra.  The background was
modeled in this case, rather than subtracted, and the model included
in the fit to the unbinned, unsubtracted source spectrum.  The results
for all three background subtractions gave fitted Fe/Si abundance
ratios that are enhanced over the solar values by factors of 10-20 by
number, or factors of 20-40 by mass.  The right panel of Figure 4
shows the fit giving the highest Fe/Si ratio.  Note the absence of
recognizable emission from S or Ar, and the weakness of the Si
emission; the emission lines at energies above 7 keV are of H-like Fe.
These results demonstrate that there is emission in Cas A that is
plausibly associated with pure Fe.

For a subset of 700 regions with the strongest Fe line emission and
the poorest fits in the Fe K band,
we tested two options for the added component representing either
shocked circumstellar medium or pure Fe and Ni ejecta.  For the
former, we take a plane-parallel shock with CSM abundances and the
average temperature (2.2 keV) and ionization age ($3\times10^{11}$
cm$^{-3}$s) obtained for the forward shock regions, all as described
above.  For the latter, we found that a simple nonequilibrium
ionization component, with a single value of the ionization age was
sufficient, with the temperature fixed at 1.95 keV and the ionization
age fixed at $8\times 10^{11}$ cm$^{-3}$s.  These values are
representative of the most enriched Fe ejecta spectra, such as those
studied by \citet{hwang03}; the fits, however, are not strongly
sensitive to the exact values used so long as they are of this
approximate magnitude.  We fix the parameters of the second component
whenever possible because the limited spectral resolution makes it
very difficult, even at these exceptionally high levels of
signal-to-noise, to constrain both model components independently.  As
before, errors were calculated for key parameters to help ensure
finding the true minimum in $\chi^2$.  Only one of these 700
ejecta+CSM fits gave $\chi^2\leq1$ in the 6-7 keV Fe K region, leading
us to conclude that the presence of a forward shock component can not
explain the strong Fe K emission in these spectra.  On the other hand,
a majority
of the {\it vpshock}+NEI fits did give acceptable $\chi^2$ values in
the 6-7 band, although the improvement was sometimes due to a better
match to the continuum rather than the line emission.  An example is
shown in the final two panel rows of Figure 5, where the same spectrum is
shown with the original single-component {\it vpshock} fit, a two
component {\it vpshock} fit for normal ejecta and shocked CSM, and
finally a two ejecta-component fit where the added ejecta component is
taken to be pure Fe (and Ni).

On a broader scale, we identify 2982 spectra where a composite ejecta
model providing additional Fe emission may be needed, based on a
fitted Fe abundance for the basic single-component model that is above
0.3 solar.  A composite model is implemented in cases where the
$\chi^2$ per degree of freedom $\nu $ for the single-component model
is greater than a user defined threshold.  We focus hereafter only on
two component models where the second (NEI) ejecta component
corresponds to pure Fe.  The three panels of Figure 6 show examples of
the Fe mass distribution in the second, ``pure'', Fe component for
values of the threshold for $\chi^2/\nu$ of 1.1, 1.2, and 1.3.
In the idealized case that fit errors are normally distributed, the
model linear, and variances known such that $\nu$ is well defined
\citep{andrae10}, these values would correspond to probabilites of
0.1, 0.01, and 0.001, respectively, that the correct model had been
fitted. Independently of the selected threshold value, the spatial
distribution of the added Fe distribution is seen to retain its main
morphological features.

\section{SNR Models}


In this section we provide an overview of our models for the evolution
of Cas A.  These are treated in a manner similar to that in
\citet{laming03} and \citet{hwang03}, for a remnant with an
  ejecta density profile with a constant core and a n=10 power-law
  envelope (after Matzner \& McKee 1999), expanding into a circumstellar
wind (density $\propto r^{-2}$).  We discuss modifications here.

We modify the models to accommodate SNR expansion into a stellar wind
``bubble'' as presented and discussed in \citet{hwang09}. The chief
motivation for this involves the interpretation of infra-red light
echoes associated with Cas A as reprocessed shock breakout radiation
\citep{dwek08}. This radiation is concentrated in the UV-EUV spectral
region, and would be strongly absorbed in photoionizing any
intervening neutral gas, which presumably would be a relic of the red
supergiant presupernova stellar wind.  Most of this opacity would lie
close to the progenitor, and so a small ``bubble'' in the
circumstellar medium could allow the breakout radiation to escape and
illuminate the surrounding dust.  This is especially true if the
exploding progenitor were compact, and the duration of breakout
radiation short, as a longer shock breakout would probably produce
sufficient photoionization to burn its way through the surrounding
wind. We note that the IIb prototype SN 1993J appears to have had
dense wind very close to the progenitor since it was detected in
X-rays a few days after explosion \citep{fransson96}.  While Cas A
resembles SN 1993J in many other respects, it would appear to differ
on this point.

Another update corrects the motion of the reverse shock\footnote{Some
  other minor modifications and corrections to Appendix A in Laming \&
  Hwang (2003) are collected in Appendix A to this paper}. In
previous work we have used a model of the ejecta where a uniform
density core is surrounded by a power law envelope. The evolution
while the reverse shock is in the outer envelope is given by
\citet{chevalier82}; this solution is then coupled to the known
asymptotic behavior of the Sedov-Taylor solution. In this phase, we
have previously held the reverse shock speed constant at its value at
the envelope-core transition, although it is known that some small
acceleration should occur during the core propagation. We now
implement a reverse shock acceleration in this phase of evolution
following the results given for a suite of models by
\citet{patnaude09}. By matching to their results, we write for the
reverse shock speed and radius:

\begin{eqnarray}
v_{r}&={3-s\over n-3}{v_b\over l_{ED}} + {n-6\over 43}x_0t\\
R_r&\!=\!\!\left[{R_b\over l_{ED}t_{core}}\!-\!{3-s\over n-3}{v_b\over l_{ED}}\ln\!{t\over t_{core}}\!-\!{n-6\over 43}x_0\!\left(t-t_{core}\right)\right]\!\!t,
\end{eqnarray}
where $v_r = R_r/t - dR_r/dt$.  Here $s$ is the power law index for
the CSM density profile, $n$ that for the ejecta envelope, $R_b$ and
$v_b$ the forward shock radius and speed, respectively, and $l_{ED}$ the
``lead factor'' (the ratio of forward to reverse shock radii,
$R_b/R_r$).  The quantity $x_0=\left(40.74M_{ej}/\rho
R_b^2\right)^{1/\left(3-s\right)}$, with $\rho$ being the density of
the CSM at the forward shock in H atoms (or equivalent mass) per
cm$^3$, $t_{core}$ is the time in years when the reverse shock hits
the ejecta core, and $M_{ej}$ is the ejecta mass in solar masses.

In Table 1 we give various models with the range of plausible ejecta
masses 2-4 M$_\odot$, all designed to match as far as possible the
known dynamics of Cas A. We take a distance of 3.4 kpc \citep{reed95}
and an explosion date of $1671.3\pm 0.9$ as determined from the proper
motions of 17 high velocity outer ejecta knots \citep{thorstensen01}.
These knots are identified on plates taken as long ago as 1951,
providing the longest time baseline for such studies, and are assumed
to be undecelerated. More recently, \citet{fesen06} have determined a
best explosion date of $1681\pm 19$ from a sample of 126 relatively
undecelerated knots located along the NW limb, observed over a time
baseline of 9 months from March to December 2004.  With 1671 as the
earliest possible explosion date, the remnant age may be up to 333
years in 2004.

Further constraints on the model are provided by the observed shock
velocities and radii.  A number of authors have made measurements of
the expansion rate of Cas A, both for its bright ejecta ring
\citep{vink98,koralesky98,delaney04} and the forward shock
\citep{delaney03}.  The most recent summary is given by
\citet{patnaude09}, who incorporate measurements over the longest time
baseline (2000--2007). They find that the expansion is slightly slower
in the N and NW filaments, consistent with the known presence of
denser CSM on the western limb of Cas A. We take the average expansion
0.31'' yr $^{-1}$ from other parts of the remnant, which (at a
distance of 3.4 kpc) gives a forward shock velocity of 5050 km
s$^{-1}$. Including all measurements gives a slightly lower average
expansion of 0.30'' yr$^{-1}$, and an inferred forward shock velocity
of 4850 km s$^{-1}$.

The forward shock radius is measured from the wispy structures seen in
4-6 keV continuum emission.  \citet{gotthelf01} use data taken in 2000
and determine an average radius for the northwest sector of $153\pm
12$'', which translates to $2.52\pm 0.2$ pc, but the true shock radius
should probably be determined from the outermost of these of these
structures rather than the average. \citet{helder08} perform a
deconvolution for various azimuthal angle ranges and give a slightly
larger forward shock radius of 160'' (2.637 pc).  There is also a
small variation in the radius around the limb.  We take a value of 2.6
pc for the forward shock radius in 2004 when the Chandra VLP
observation was taken.

The reverse shock radius is harder to quantify. \citet{gotthelf01} and
\citet{helder08} give a range of radii 1.52 -- 1.73 pc, depending on
location in the remnant. \citet{morse04} measure the reverse shock
velocity from proper motions derived from two Hubble Space Telescope
WFPC2 images separated by two years. At a reverse shock radius of
106'' -- 122'' (1.75 -- 2.01 pc) the reverse shock velocity with
respect to the expanding unshocked ejecta is of order $\sim 2000$
km s$^{-1}$.

The close match of Cas A's forward shock expansion parameter of 0.66
to the Sedov-Taylor value for expansion into an inverse square density
profile has focussed interest in low ejecta mass models that allow
faster evolution to the Sedov-Taylor limit. Even so, models generally
predict too high a shock velocity and/or too small a radius for the
forward shock compared to observations. To increase the forward shock
radius, it has been suggested \citep[][and see above]{hwang09}
that Cas A exploded into a small bubble
in the circumstellar medium that developed with a transition from a
slow dense red supergiant wind to a fast tenuous wind shortly before
explosion.  Another possibility is that cosmic ray
energy losses at the forward shock provide extra deceleration to
suffciently reduce the expansion parameter \citep{patnaude09}.

The expansion parameter for a radiative blast wave can be different
from the Sedov-Taylor limit in the case that radiative losses from the
shock are significant, or if the shock dissipates energy via cosmic
ray acceleration.  The two cases may be treated in a similar
manner. Following \citet{liang00} and generalizing to the case of a
stellar wind preshock density profile $\rho\propto r^{-s}$, the radius
of a radiative blast varies with time as
\begin{equation}
r\propto t^{1/\left[4-s-\left(3-s\right)\alpha\right]},
\end{equation}
where $\gamma$ is the adiabatic index of the gas,
\begin{eqnarray*}
\alpha=\left\{2-\gamma+\sqrt{\left(2-\gamma\right)^2 +
  4\left(\gamma_1-1\right)}\right\}/4, 
\end{eqnarray*}
and
\begin{eqnarray*}
\gamma_1=\left\{\sqrt{1+\epsilon\left(\gamma^2-1\right)}-1\right\}\times
2/\left(\gamma-1\right)\epsilon -1 
\end{eqnarray*}
is the adiabatic index modified by
the loss of a fraction $\epsilon$ of the shock energy to radiative (or
cosmic ray) losses. If $\epsilon = 0$ so that $\gamma =\gamma _1$,
then $\alpha=1/2$ for all $\gamma$ and $r\propto
t^{2/\left(5-s\right)}$.  Note that energy going into cosmic rays that
remain trapped with their energy in the shock does not change the
dynamics, except through modifying $\gamma$. If all the shock energy
is radiated ($\epsilon =1$), then $\alpha = 1-\gamma /2$.  Taking
$\gamma =5/3$, the expansion parameter changes from 0.4 to 0.28 for
$s=0$, and from 0.67 to 0.55 for $s=2$. Consequently, rather small
deviations from the unmodified expansion parameter are expected for
more modest cosmic ray energy losses. Cosmic ray energy losses of 10\%
only change the $s=2$ expansion parameter from 0.67 to 0.65,
decreasing to 0.63 for 30\% losses. \citet{patnaude09} find variations
in the expansion parameter of this order. \citet{bedogni84} give full
self-similar solutions for several such cases.

The plasma density ahead of the forward shock is derived by
\citet{willingale03} from fits to XMM data. The baryon mass and
filling factor quoted in their Table 3 give a preshock density of
hydrogen atoms or equivalent mass of 1.47 amu cm$^{-3}$,
assuming a shock compression of a factor of 4. Their quoted value for
the mass of shocked CSM, 8.31 $M_{\sun}$, combined with the dimensions
outlined above gives a larger preshock density of 1.99
cm$^{-3}$. These values give the time invariant quantity $\rho
r_f^2=10 - 13.5$ cm$^{-3}$pc$^2$ where the density $\rho$ at the
forward shock is in H atoms (or equivalent mass) cm$^{-3}$ and the
forward shock radius $r_f$ is in pc.

Similar results are obtained from the fits of thermal bremsstrahlung
spectra to Suzaku data \citep{maeda09}, interpreted as emission from
shocked circumstellar medium.  They fit the X-ray continuum with a
variety of composite models.  The preshock ion number density derived
from their thermal bremsstrahlung and power law fit is
$2.1/\left<Z^2\right>$ cm$^{-3}$, where $\left<Z^2\right> = \sum
_in_iZ_i^2/\sum _in_iZ_i$ is an average charge per ion coming from the
thermal bremsstrahlung emissivity. For gas composed of equal number
densities of H and He, $\left<Z^2\right> = 1.7$ and the preshock
density is 1.24 cm$^{-3}$ in H atoms of equivalent mass. Taking fit
results using thermal bremsstrahlung plus SRCut\footnote{The SRCut
  model is a widely available model that provides the most maximally
  curved nonthermal spectrum that is plausible, i.e., it imposes an
  exponential cutoff on the power-law energy spectrum of the
  synchrotron emitting electrons \citep{reynolds99}.} or a cut-off
power law gives a higher density of 1.5 cm$^{-3}$ and $\rho r_f^2=8.4
- 10$ cm$^{-3}$pc$^2$.


\section{Discussion}

\subsection{The Elemental Composition of the Cas A Ejecta}
In previous works \citep{laming03,hwang03} we have discussed the use
of models of SNR evolution to interpret the fitted ionization age of
ejecta spectra in terms of the time elapsed since reverse shock
passage, hence allowing the inference of a Lagrangian mass
coordinate. This is successful in a limited number of carefully
selected regions of Cas A, but is evidently not the case in general.
Figure 2 shows the ionization age value strongly peaked near
$\log\left(n_et\right)\simeq 11.3$. Converting ionization age to mass
coordinate, the implication is that all the observed ejecta are piled
up at a mass coordinate of 0.2 (where 0 marks the center of the ejecta
and 1.0 the outermost extent), instead of being uniformly distributed
with mass coordinate. It appears that the ionization age is not an
accurate indicator of ejecta mass coordinate.  We suggest that this
must arise from the interaction of the reverse shocked ejecta with
secondary shocks propagating within the SNR shell, which ``refresh''
the ionization age. These secondary shocks are a direct
  consequence of density inhomogeneities.  The forward and reverse
  shocks that encounter them are both transmitted through the density
  structure and also reflected back into previously shocked gas. Such
a scenario has previously been considered \citep{laming01a,laming01b}
as a possibility for electron acceleration with a view to explaining
the hard X-ray emission of Cas A. More recently, \citet{inoue10}
consider a similar model for ion acceleration at shocks.

Taking a typical electron density of 200 cm$^{-3}$ \citep[][and see
  below]{lazendic06}, an ionization age of $2\times 10^{11}$
cm$^{-3}$s implies a shock interaction around 30 years prior to the
observation. Given this ``obscuration'' of the ionization age
determined by the reverse shock, we estimate the density in each fit
region from the emission measure assuming
\begin{equation}
EM = \sum _in_in_eZ_i^2V = n_eV\sum_in_iZ_i^2 =n_e^2V\left<Z^2\right>
\end{equation}
where $\left<Z^2\right>= \sum _in_iZ_i^2/\sum _in_iZ_i=\sum
_in_iZ_i^2/n_e$, and the sum $i$ is over all ions in the plasma.  This
generalizes the usual definition for a hydrogen dominated plasma to
the case of heavy element rich supernova ejecta, where the continuum
emission is dominated by thermal bremsstrahlung emission going as
$Z_i^2$. The plasma volume is $V$, and the plasma electron density
then follows with an appropriate assumption about $V$. The simplest
assumption is that $V=A^{3/2}$ (where $A$ is the area on the sky of the
spectral extraction region), and we modify this to $V=A^{3/2}\times f$ where
$f$ is the filling factor. The filling factor is a needed correction
for unresolved density structure in the ejecta, particularly for SNR
evolution into a stellar wind ($\rho \propto r^{-2}$), as a strong
density spike is expected in the ejecta density profile.  As discussed
in Appendix B, we characterize the filling factor $f$ by the density
scale length $L_p$, where $f=L_p/\sqrt{A}$, and derive an approximate
value for $L_p$ of $2.5''$ for a distance to Cas A of 3.4 kpc.

The ejecta mass in each imaged region is then estimated as
$EM/n_e/\left<Z^2\right> \times \sum _in_im_i/n_e$, from which the
masses of individual elements are obtained as fractions of the total
ejecta mass from the fitted element abundances. In Table 2 we give the
masses of O, Ne, Mg, Si, S, Ar, and Fe resulting from our fits, for
assumed filling factors corresponding to a density spike thickness of
2.5'' and 5''; the latter arises as one observes the SNR ejecta shell
in both front and back along the same line of sight.  In this latter
case, the complete X-ray emitting ejecta mass inferred is 2.84
$M_{\sun}$, and the total ejecta mass including unshocked ejecta is
3.14 $M_{\sun}$ with reference to Table 1; this gives a value of $\rho
r_f^2$ from Table 1 at the upper end of values inferred elsewhere as
discussed in section 3.1.  Extrapolation to other cases is
simple, since element masses are proportional to $\sqrt{f}$. Table 2
also gives plane of the sky velocities for each element (calculated
from the position of the center of mass of each element, and assuming
homologous expansion), and for the SNR as a whole, and compares
abundance ratios relative to O with solar system values.

We also comment here that Cas A is thought to have retained some H at
the time of explosion \citep{fesen01,fesen91,chevalier03}, and this
may contribute to the thermal bremsstrahlung emission here attributed
to O. In this case, our ejecta mass estimate would be an
underestimate, and if the H (and presumably also He) were not
uniformly distributed, the recoil velocity we attribute to the O
ejecta could be in error. It is difficult to assess an uncertainty
here, but we note that the amount of light elements involved is likely
rather small and unlikely to alter our conclusion of an approximately
3 $M_{\sun}$ ejecta mass.

In Table 3 we give both the mass of observed Fe ejecta associated with
incomplete Si-burning, and that associated with the ``pure'' Fe
component which can arise either from complete Si-burning or
$\alpha$-rich freeze out.  We list the results for a range of initial
$\chi^2/\nu$ for the single {\it vpshock} fit, as discussed in section
2.4.  We also consider the the quantity ${1\over\Gamma\left(\nu
  /2\right)}\int _{\chi ^2/2}^{\infty}x^{\nu /2-1}\exp -xdx$.  In the
case that the fit parameters are normally distributed, and if
additional conditions are met as discussed at the end of susbsection
2.4, this quantity would be the probability for degrees of freedom
$\nu$ that the single component fit used the correct model.  The last
column column of the table gives the fraction $f_{reg}$ of the 2982
ejecta spectra considered that were fitted with the composite {\it
  vpshock + NEI} model based on the threshold $\chi^2/\nu$ cutoff for
the single component model.  As can be seen from the table, demanding
a higher threshold for $\chi^2/\nu$ (or lower estimated probability)
to implement the second ``pure'' Fe fit component reduces
Fe$_{\alpha}$. At a cutoff value of $\chi^2/\nu=1.2$ corresponding to
an estimated probability of 0.01, the fraction $f_{reg}$ begins to be
much greater than the estimated probability, arguing that we are
seeing a real physical effect.

\subsection{The Mass and Distribution of Fe}

Our analysis of more than 4300 ejecta regions gives a mass of X-ray
emitting Fe in the Cas A ejecta of $\sim 0.09 - 0.13 M_{\sun}$,
depending on the assumed filling factor and threshold value of reduced
$\chi^2$ for acceptance of the ``pure'' Fe component. There are
uncertainties in this estimate, as it is conceivable that the Fe
ejecta are clumped in a different manner than the other ejecta,
possibly as in the simulations of Hungerford et al. (2005), and thus
require a different filling factor than we have used.  If, for example
the filling factor for Fe should be smaller, this would yield a
smaller Fe mass.  The Fe mass is comparable to that expected
\citep{eriksen09}, but gives only the Fe detected in the reverse
shocked X-ray emitting ejecta.  According to our models, an extra 0.18
- 0.3 $M_{\sun}$ of ejecta may be unshocked, interior to the reverse
shock, and visible primarily in the infra-red.

Emission at the center
of the remnant that is believed to be from unshocked ejecta is seen in
infrared {\em Spitzer} observations, but the emission is of [Si II].  As
discussed by DeLaney et al (2010), it is coincident with free-free
absorption seen in the radio by Kassim et al. (1995), and appears to
correspond to cool ($< 1000$ K), low density gas that has been
photoionized.  DeLaney et al. conclude, however, that it is unlikely
that a significant fraction of these unshocked ejecta could be Fe.
Relatively little infra-red Fe emission is observed at all in the
remnant, and the main [Fe II] line at 26 $\mu$ is blended with [O IV].
\citet{isensee10} demonstrate that [O IV] is more plausibly the
dominant component of this blend, though they do claim a 2$\sigma$
detection of [Fe II] 17.94 $\mu$m from the center when all spatial
bins are summed.  Unblended [Fe II] emission at 17.94 $\mu$m
\citep{ennis06} and 1.64 $\mu$m \citep{rho03} is more convincingly
detected, but appears to be associated primarily with the bright
ejecta ring, and not the interior.  [Fe II] lines are also absent or
very weak in optical and near infra-red observations of Cas A ejecta
knots. \citet{hurford96} only report marginal detections of [Fe II]
8617 \AA\ in spectra of fast moving knots (FMKs), and
\citet{gerardy01} report similarly on the [Fe II] lines between 1 and
2 $\mu$m in FMKs, although these transitions are easily detected in
the spectra of quasi-stationary flocculi (QSFs). \citet{eriksen09}
detect [Fe II] 1.257 $\mu$m and 1.644 $\mu$m lines from 15 FMKs and 4
QSFs, and use these to estimate the reddening.

Further consideration of the dust in Cas A sheds light on the possible
presence of unshocked Fe. Large masses of cold dust had been reported
in Cas A from submillimeter observations \citep{dunne03}, but it has
been pointed out that some of the detected emission originates from
foreground molecular clouds \citep{krause04, wilson05}. This emission
can also be explained with $<10^{-3}$ M$_\odot$ of conducting dust
needles (i.e., Fe) that are formed in the ejecta (Dwek
2004). \citet{rho08} favor FeO dust to explain the {\em Spitzer} infrared
spectra and conclude that Fe dust may be present at masses up to
$10^{-2}M_{\sun}$.  

More recently, \citet{nozawa10} suggest the presence of significant
quantities of cool dust, a conclusion that is supported by recent
infrared observations with Herschel \citep{barlow10}, AKARI and BLAST
\citep{sibthorpe10}.  The cool dust has a temperature of about 35 K
based on the infrared flux densities, and is presumably unshocked
ejecta as it is confined to the central regions of the remnant.  The
emission is consistent with a silicate dust composition, and the
inferred dust mass is about 0.06-0.075 M$_\odot$.  Nozawa et
al. (2010)'s calculations are for dust formation in a Type IIb event
(with an eye toward Cas A) and indicate that little dust would be
associated with the innermost Fe-Ni layer due to the extended
radioactive heating: the gas density drops too low before the
temperatures are low enough for Fe or Ni grains to condense. Up to
about $10^{-3} M_{\sun}$ of Fe could be locked up in FeS grains, but
highly Fe-rich ejecta would not necessarily have a dust signature, as
appears to be borne out by the infrared
observations. \citet{cherchneff10} take a chemical kinetic approach
and find rather more FeS grain formation in a 20 $M_{\sun}$ model SN
with unmixed ejecta (for Cas A, the most appropriate of the cases they
consider), with about 0.021 $M_{\sun}$ of Fe in FeS.  A mass this high
is unlikely for Cas A, however, given that its highly stripped
progenitor underwent a Type IIb event.  Dust will form relatively more
efficiently in a normal Type II event because the overlying stellar
envelope restricts ejecta expansion and preserves higher densities
favorable for dust condensation.

While Fe appears to be scarce both in the shocked ejecta and the
unshocked infrared emitting ejecta at the center of the SNR, Fe may
yet be present in the unshocked ejecta given that the intrinsic
emissivities of the infra-red lines of Fe II are rather weak compared
to, say, Si II.  Another complication is that there may be a
population of low-density Fe ejecta, whose composition and
distribution are currently unconstrained.  Sufficiently large clumps
of Fe might be expected to reside in low density regions due to
inflation by radioactivity.  Eriksen (2009) note that the lack of
detectable Fe lines in the {\em Spitzer} data indicate that any Fe
material still interior to the reverse shock must indeed be at a low
density, though they also suggest that the absence of Fe is more
likely to be an abundance effect. While an accurate assessment of
infra-red limits must be deferred to another work, strong outward
mixing of Fe into at least the He layer has been inferred for the Type
IIb prototype SN 1993J based on modelling its light curve
\citep{shigeyama94}.  Cas A resembles SN 1993J, not only in its
explosion spectrum (seen as a light echo, Krause et al. 2004), but in
having a similar and low pre-explosion main sequence mass, and a
binary companion implicated by its strong pre-supernova mass loss.

If our tentative conclusions about the paucity of Fe in the unshocked
ejecta survive more detailed scrutiny, then almost all of the Fe
ejected by the supernova is now well outside the reverse shock and
visible in X-rays, with very little left in the center of the
remnant. This would be true even for the ``pure'' Fe component, which
could be the ashes of $\alpha$-rich freezeout. The inference that the
Cas A neutron star has a carbon atmosphere \citep{ho09} would seem to
support the notion that the interior ejecta composition is dominated
by elements lighter than Fe.  Along these lines, recent work by
\citet{ouyed11} raises the interesting possibility that under the
right conditions, a quark nova of the incipient neutron star could
produce $^{44}$Ti by spallation at the expense of $^{56}$Ni, and also
cause C formation consistent with the inferred composition of the
compact object in Cas A.\footnote{In this scenario, the compact object
  is predicted to be a radio-quiet quark star surrounded by material
  rich in C.}

Closely related to explosive Fe nucleosynthesis is the production of
$^{44}$Ti.  Cas A is one of only a handful of supernova remnants
showing clear evidence for $^{44}$ Ti in the ejecta.  An initial mass
of $^{44}$Ti of $1.6\times 10^{-4} M_{\sun}$ has been inferred for Cas
A from observations with the Compton Gamma Ray Observatory/COMPTEL
\citep{iyudin94}, BeppoSAX and the INTEGRAL IBIS/ISGRI instrument
\citep{vink01,renaud06}. Based on spherically symmetric simulations,
designed to match the Cas A progenitor \citep{young06,eriksen09},
\citet{magkotsios10} predict $\sim 10^{-4} M_{\sun}$ of $^{44}$Ti, and
0.25 $M_{\sun}$ of $^{56}$Ni. These are both within a factor of 1.6 of
measured values (taking the 5'' filling factor in Table 2), but are
discrepant in opposite directions so that the mass ratio
$^{56}$Ni$/^{44}$Ti$\simeq 2500$ compared with the observed ratio of
$\sim 10^3$.  A similar outcome of the earlier calculations of
\citet{the98} and others \citep[reviewed in][]{the06} led
\citet{nagataki98} to suggest that this ratio may be reduced in an
asymmetrical explosion. Another way that $^{44}$Ti production could be
enhanced relative to Fe is a quark nova, as discussed above
\citep{ouyed11}.  

The results of \citet{magkotsios10} also enable an estimate of the Fe
expected from $\alpha$-rich freeze out compared with that formed in
incomplete Si-burning. Their Figure 17 suggests that approximately
half the Fe should be associated with a mass ratio
$^{56}$Fe$/^{28}$Si$ > 10$ and half with $^{56}$Fe$/^{28}$Si$ <
10$. If we take a $\chi^2/\nu$ cutoff of 1.2 from our Table 3, we
infer that approximately 1/3 of the Fe in Cas A could have formed in
$\alpha$-rich freeze out and thus be in regions with very small
$^{28}$Si masses, in qualitative agreement with the model estimate.
In Cas A, the ``pure'' Fe is placed at least 0.75 $M_{\sun}$ out from
the center in 3 $M_{\sun}$ of ejecta by the the characteristic
ionization age of $> 2\times 10^{11}$ cm$^{-3}$s. According to
\citet{magkotsios10}, this is well outside the ejecta region where the
Fe was formed, and into the outer layers where the composition would
originally have been dominated by O, Si, C, and Ne. The fact that
essentially all of the Fe so far is found at such locations, and not
in the center of the remnant inside the reverse shock suggests the
operation of a strong instability, similar to that inferred in SN
1993J \citep{shigeyama94}.

Key information about the explosion of Cas A will be provided by the
distribution of $^{44}$Ti, and this will be revealed by observations
with NUSTAR \citep{harrison10}.  If a quark nova did take place during
the supernova explosion that produced Cas A, the $^{44}$Ti
distribution should have a strong central component.  Unshocked Fe in
the center of the remnant will also contribute to such a component.
In the absence of a quark nova, the $^{44}$Ti is expected to follow
that of the pure Fe, which we observe so far in X-rays well outside
the reverse shock. It is interesting to note that in the case of SN
1987A, the ejecta are currently heated by the decay chain of
$^{44}$Ti, which appears to be located in the center of the remnant,
with ballistic expansion velocity of order 3000 km s$^{-1}$
\citep{kjaer10}. The youngest Galactic remnant, G1.9+0.3
\citep{borkowski10}, is more like Cas A in this respect, in that most
of its $^{44}$Ti in an X-ray bright region beyond the reverse shock,
and a smaller portion unshocked interior. Considering the young age
and fast expansion of G1.9+0.3, this may mean that $^{44}$Ti has been
mixed out even further into the ejecta than is the case in Cas A,
although here the SN origin of G1.9+0.3 is unknown.

Some insight into the motion of the ``pure'' Fe component can be
gained from the motion of $^{44}$Ti observed with INTEGRAL
\citep{martin08,martin09}. The SPI instrument measures the width and
the shift of the 1157 keV decay line of $^{44}$Ca, which is the final
step in the chain of decays from $^{44}$Ti.  It is found that the
$^{44}$Ti is receding from the observer with a velocity of about 500
km s$^{-1}$, and has an intrinsic width, or velocity dispersion, of
430 km s$^{-1}$. Since the expansion velocity of the Fe is close to
4000 km s$^{-1}$, if the Fe is associated with the $^{44}$Ti, it must
be expanding in a direction close to the plane of the sky. The
velocity dispersion indicates that the $^{44}$Ti presumably occupies a
volume significantly smaller than that outlined by the ``pure'' Fe,
with a half angle subtended at the explosion center of order $\sin
^{-1}\left(430/4000\right)\simeq 6$ degrees.

\subsection{The Neutron Star Kick}
The neutron star is known to be recoiling from the explosion center
inferred from the optical knot motions with a velocity in the plane of
the sky of 330 km s$^{-1}$ \citep{thorstensen01}. The direction of
recoil is also nearly perpendicular from the NE-SW axis established by
the jet and counter-jet, which presumably delineate an axis of
rotation in the progenitor. This makes a kick arising from the action
of a standing accretion shock instability appealing. It is known, at
least in the limit of slow rotation, where distortion of the shock
front from spherical symmetry can be neglected, that the instability
grows fastest in the prograde spiral mode
\citep[e.g.][]{blondin07,yamazaki08,iwakami09}, i.e., perpendicular to
the symmetry axis. The pulsar kick imparted by the second supernova
that formed the double pulsar system PSR B1913+16 is also inferred to
have been largely perpendicular to the progenitor spin axis
\citep{wex00}. The magnitude of the observed recoil velocity is also
well within the accessible range generated by numerical simulations of
such instabilities, albeit these are so far for initially nonrotating
supernova cores \citep{wongwat10,nordhaus10}. At higher rotation
rates, it appears that the instability is suppressed
\citep{burrows07}, though the deformation of the accretion shock by
rotation may seed an axisymmetric quadrupole oscillation
\citep{iwakami09}.

In the case that the kick involves instability at the accretion shock,
following the suspicion advanced by \citet{wongwat10}, we would expect
the Fe ejecta to recoil predominantly in the opposite direction to the
neutron star. In Cas A, this appears not to be the case. Most of the
Fe, and especially most of the ``pure'' Fe, is recoiling towards the
east, within 90 degrees of the neutron star motion. However, the
remnant as a whole, whose mass is dominated by O, is actually
recoiling to the north, close to 150 degrees from the neutron star
motion, with a plane of the sky velocity of about 700 km s$^{-1}$,
which perhaps suggests a hydrodynamic origin for the kick.
\citet{wongwat10} discuss a 15 $M_{\sun}$ explosion, whereas the Cas A
progenitor was considerably less massive upon explosion such that
reverse shocks and associated nucleosynthesis during the explosion are
less likely to occur. Our estimate of the remnant recoil assumes
homologous expansion, and the approximations made in assessing the
ejecta mass have been detailed in sections 3 and 4.1.

\section{Conclusions}

We consider our efforts here to be a {\em first} attempt at a
comprehensive view of the X-ray emitting ejecta in Cas A on few
arcsecond angular scales.  We have carried out at considerable effort
the most detailed data analysis that was practicable for us, but there
is ample room for improvement.  The most obvious shortcoming of this
work is the need for a more sophisticated definition of spectral
extraction regions, and hand-in-hand with that, yet more extraction
regions.  In previous work, we had individually chosen regions to
correspond to identifiable knotty or filamentary features.  Here we
were obliged for the sake of (relative) expediency to instead use
regions that, while adjusted crudely for overall surface brightness,
have been defined without reference to coherent structures in the
surface brightness.  Full use of the Chandra angular resolution in
fact requires a more sophisticated definition of the extraction
regions.  The main advantage to be obtained, however, is that this may
allow the use of simpler NEI models with a single ionization age,
rather than parallel shock models with a linear distribution of
ionization ages, and thus a much clearer connection with analytical
hydrodynamical models \citep[see for example,][]{laming03,hwang03}.
It may also eliminate the requirement for the filling factor
correction used here.  Finally, the analysis would also benefit from a
more sophisticated treatment of the spectral background.

We have arrived at a favored ejecta mass of about 3 $M_{\sun}$ based
both on the observed SNR dynamics, and on an accounting of the X-ray
emission and element abundances throughout the remnant.  The increased
signal/noise afforded by the long exposure time, and by the relatively
large size of extraction regions has allowed us to identify regions of
the remnant where the Fe ejecta is to a large extent ``pure'', in that
the mass ratio Fe/Si $>$ 10.  We seek to interpret this ejecta
component as alpha-rich freeze out ashes, where the recently
discovered \citep{iyudin94,vink01,renaud06} $^{44}$Ti and its decay
products likely reside. Essentially all the Fe detected so far, and
presumably also the $^{44}$Ti and its decay products, are not located
in the center of the remnant but have been ejected well out into the
overlying ejecta, probably by the action of hydrodynamic instabilities
during the explosion. Such outward mixing of $^{56}$Ni has been seen
in the Type IIb prototype SN 1993J \citep{shigeyama94}.  With respect
to the distribution of $^{44}$Ti, SN 1987A presents a stark contrast to
Cas A, with ejecta containing $^{44}$Ti located at the center with a
characteristic expansion speed of $\sim 3000$ km s$^{-1}$
\citep{kjaer10}.  G1.9+0.3 may represent an intermediate case
\citep{borkowski10}.

Finally, Cas A has a well documented displacement of the neutron star
formed in the explosion \citep{thorstensen01}, which is close to
perpendicular to the NE-SW ``jet''-axis. While contrary to
expectations \citep[e.g.][]{wongwat10}, the Fe in Cas A does not
recoil in the opposite direction to the neutron star, as far as we can
tell, but the remnant as a whole does. We take this as reinforcing the
notion that neutron star kicks in core-collapse supernova explosions
may derive from purely hydrodynamic mechanisms, as recently suggested
\citep[e.g.][]{nordhaus10}. Together with motion and location of the
Fe ejecta, and the likelihood that what is observed as a NE-SW ``jet''
really is an artefact of an asymmetric explosion, we speculate that an
instability of the standing accretion shock (SASI) in a rotating
progenitor is the most promising model for understanding the explosion
that formed Cas A.

\acknowledgements We thank Roger Chevalier, Elisabetta Micelotta, Rob
Petre, and the referee for helpful comments on the paper. We gratefully
acknowledge support through NASA grants to the Chandra Guest Observer
Program and NNH04ZSS001N to the Long Term Space Astrophysics Program.
JML was also supported by basic research funds of the Office of Naval
Research.

\appendix
\section{Revisions to Appendix A in Laming \& Hwang (2003)}
We collect here a few minor corrections and modifications to the text in Appendix A of \citet{laming03}.
Equations A1 and A2 should read
\begin{equation}
t_0=473.6 M_{ej}^{5/6}E_{51}^{-1/2}\rho ^{-1/3} {\rm yr},
\end{equation}
\begin{equation}
x_0=3.43 M_{ej}^{1/3}\rho ^{-1/3} {\rm pc.}
\end{equation}

The expressions for the lead factor and the pressure ratio between forward and reverse shocks following equation A5 are now more accurately given by
\begin{equation}
l_{\rm ED}=1.0+ {8\over n^2} +{0.4\over 4-s}
\end{equation}
\begin{equation}
\phi _{\rm ED}=\left(0.65 - \exp\left(-n/4\right)\right)\sqrt{1-s/8}.
\end{equation}

Finally, as pointed out to us by Elisabetta Micelotta, some of the discussion following
equation A8 in \citet{laming03} requires clarification. Equation A5 can be derived from the envelope form of equation A8, i.e. from equations A7 and A8 in \citet{truelove99}. This procedure requires $v_{ej}=\sqrt{\left(10/3\right)
\left(n-5\right)/\left(n-3\right)}$ going from equation 74 to 75 in \citet{truelove99}. In their equation 31, \citet{truelove99} give $v_{core}=\sqrt{\left(10/3\right)
\left(n-5\right)/\left(n-3\right)}$, which implies $w_{core}=v_{core}/v_{ej}=1$. However
their derivation of equation 31 assumes $w_{core}\rightarrow 0$ in their derivation. A
``cleaner'' route to equation 75 of \citet{truelove99} is to take their values of
$v_{core}$ and the ejecta mass fraction in the core derived under the condition
$w_{core}\rightarrow 0$, and substitute into equation 3 of \citet{chevalier82}. The
condition $R_b=l_{\rm ED}v_{core}t_{core}$ still gives the time $t_{core}$ when the
reverse shock hits the ejecta core. The only other change connected with this discussion is that equation A11 for $t_{conn}$ in \citet{laming03} should depend on $v_{core}$ and not $v_{ej}$.

In actual fact, the difference between the approximations $w_{core}=0$ and $w_{core}=1$ is
not all that large; $v_{core}=\sqrt{\left(10/3\right)
\left(n-5\right)/\left(n-3\right)}\rightarrow \sqrt{2}$ in these limits, which coincide for
$n=8$. The core fraction of the ejecta mass varies between $\left(n-3\right)/n$ and 1, and only $v_{ej}$ which goes from $\infty$ to $v_{core}$ between these two limits is sensitive.

\section{The Ejecta Filling Factor}

A correction is needed for unresolved density structure in the ejecta.
Models of supernova remnants expanding into a stellar wind ($\rho\propto
r^{-2}$) density profile generally show a strong density ``spike'' in
the radial direction at the contact discontinuity, and if the reverse
shock is still propagating through the outer power law portion of the
ejecta density profile, the density at the contact discontinuity is
formally infinite \citep{chevalier82}. Later in its evolution, the
density spike is still present \citep{chevalier03}, with a width
corresponding to about 2 arcsec. This is comparable to or smaller than
the size of many of our fit regions, so we expect that ejecta
structures will not always be resolved, and a filling factor
correction is therefore needed when calculating the plasma electron
density (see subsection 4.1). We take the volume of each spectral
region o be $A^{3/2}$ where $A$ is the area imaged in arcsec$^2$. This is
modified to $A^{3/2}f$, where $f$ is the ``filling factor'' of the
ejecta density profile.

We estimate a quantitative filling factor $f$ to be applied to our fit
results in terms of the density scale length $L_{\rho}$ as
$f=L_{\rho}/\sqrt{A}.$\footnote{We are considering emission from a
  ``sheet'' of plasma, so only one dimension is compressed, rather
  than from a clumpy medium, where all three dimensions would require
  correction.}  In deriving $L_{\rho}$, we follow \citet{hamilton84},
who give a similarity solution for ejecta structure close to the
contact discontinuity. Treating the expansion of uniform density
ejecta, they use a coordinate system with $z=0$ at $r=v_{exp}t$, the
outer limit of the ejecta at time $t$ if they were freely expanding
without interacting with the reverse shock.  The location of the
reverse shock is $z_s$, giving dimensionless Eulerian and Lagrangian
coordinates $\zeta=z/z_s$ and $\zeta _0=z_0/z_s$ respectively, where
$z_0$ is the initial value of $z$ in a Lagrangian plasma element at
the time that the plasma was shocked, i.e. $z_s$ at the time of shock
passage. With these definitions, close to the contact discontinuity
they find (their equation 13)
\begin{eqnarray}
\rho &= \rho _s\left(p\over p_s\right)^{3/5}\zeta _0^{-6/5}\cr
\zeta &= V_s\left(p\over p_s\right)^{-3/5}{\zeta _0^{\nu}\over\nu }+\zeta _c
\end{eqnarray}
where $\nu = 23/15$ for a stellar wind external density profile,
$\zeta _c$ is the value of $\zeta$ at the contact discontinuity, and
$V_s=1/12$ is the dimensionless reverse shock velocity. The pressure
is $p$, and $p_s$ is the pressure at the reverse shock.  Neglecting
the dependence of $p$ on $r$ in this region, we write
\begin{equation}
{\partial\rho\over\partial r}={\partial\rho\over\partial\zeta _0}
{\partial\zeta _0\over\partial\zeta}{\partial\zeta\over\partial r}={6\over 5}{\rho
\over z_sV_s}\left(p\over p_s\right)^{3/5}\zeta _0^{-\nu }={6\over 5z_s\nu}{\rho
\over\left(\zeta -\zeta_c\right)} ={\rho \over L_{\rho}}
\end{equation}
so that the density scale length $L_{\rho}= 23\left(\zeta -\zeta
_c\right) z_s/18$. We calculate the average of this quantity
$\left<L_{\rho}\right>=\int _{r_s}^{r_c}\rho L_{\rho}r^2dr/\int
_{r_s}^{r_c}\rho r^2dr$ using
\begin{eqnarray}
\int _{r_s}^{r_c}\rho r^2dr &=\int _{r_s}^{r_c}\left(p\over p_s\right)^{3/5}{\rho _s\zeta _0^{-6/5}}r^2dr
=\rho _s\left(p\over p_s\right)^{\left(3/5-12/23\right)}\left(5z_s\over 92 \right)^{18/23}\int _{r_s}^{r_c}{r^2\left(r_c-r\right)^{-18/23}}dr\cr
\int _{r_s}^{r_c}L_{\rho}\rho r^2dr &={23z_s\over 18}\int _{r_s}^{r_c}\left(p\over p_s\right)^{3/5}\rho _s{\left(\zeta-\zeta _c\right)\over\zeta _0^{6/5}}r^2dr
=\rho _s\left(p\over p_s\right)^{\left(3/5-12/23\right)}\left(5z_s\over 92 \right)^{18/23}{23\over 18}\int _{r_s}^{r_c}{r^2\left(r_c-r\right)^{5/23}}dr.
\end{eqnarray}
The integrals are evaluated using the substitution $u=r_c-r$, with the
final result
\begin{equation}
\left<L_{\rho}\right>={23\over 18}\left\{{23\over 18}\left(1-r_s/r_c\right)^{28/23}-{46\over 51}\left(1-r_s/r_c\right)^{51/23}+
{23\over 74}\left(1-r_s/r_c\right)^{74/23}\over
{23\over 5}\left(1-r_s/r_c\right)^{5/23}-{23\over 14}\left(1-r_s/r_c\right)^{28/23}
+{23\over 51}\left(1-r_s/r_c\right)^{51/23}\right\}r_c.
\end{equation}

Including the variation of $p$, $L_{\rho}$ in equation B5 is modified
to
\begin{equation}
L_{\rho}=z_s\left(\zeta -\zeta_c\right)\left\{5\nu /3 -\partial\ln p/\partial\ln\zeta _0\over 2-\partial\ln p/\partial\ln\zeta _0\right\}.
\end{equation}
From Table 1 of \citet{hamilton84}, $\partial\ln p/\partial\ln\zeta
_0$ varies from -1 at the reverse shock to about -0.6 at the contact
discontinuity, which reduces the value for $L_{\rho}$ derived assuming
$\partial\ln p/\partial\ln\zeta _0=0$ by 5\% to 7\%. For completeness, if the external
density $\propto r^{-3}$, then $\nu = 6/5$ and
$L_{\rho}=z_s\left(\zeta -\zeta _c\right)=r_c-r$, independent of
$\partial\ln p/\partial\ln\zeta _0$. The filling factor is then
calculated as $f=L_{\rho}/\sqrt{A}$, with $L_{\rho}$ and $A$ here
expressed in arcsec and arcsec$^2$ respectively.

As we discuss in the paper, our data suggest that many secondary shocks
reflect from the blast wave and reverse shock as they encounter
density inhomogeneities in Cas A. We therefore use equation B4 with the
tacit assumption of constant pressure near the contact
discontinuity. \citet{truelove99} point out that this approximation is
also made by \citet{hamilton84} in their energy equation.  With
$r_s/r_c = 0.9$ we find $L_{\rho}= 0.021 r_c=1.26\times 10^{17}$ cm,
which at a distance of 3.4 kpc subtends an angle of 2.48 arcsec. In
this last step we are extrapolating slightly out of the formal range
of validity of the \citet{hamilton84} model.  There,
$z_c/z_s=\left(v_{exp}t-r_c\right)/\left(v_{exp}-r_st\right)=0.953$,
which requires an expansion velocity of the outermost ejecta greater
than 20,000 km s$^{-1}$ for values of $r_c=1.95$ pc and $r_s=1.7$ pc
appropriate for Cas A. It is also worth bearing in mind other caveats
to the application of this model. \citet{hamilton84} assume uniform
density ejecta, while we adopt the prescription of \citet{truelove99}
of a core-envelope density profile. We have also introduced the
circumstellar ``bubble'' into which Cas A is assumed to have exploded.

\clearpage

\begin{table}[t]
\begin{center}
\caption{Model Parameters}
\begin{tabular}{ccccccccccccc}
\tableline\tableline
$M_{ej}/M_{\sun}$& $E_{51}$& $v_b$& $R_b$& $v_r$& $R_r$& $\rho r_f^2$& $R_{bub}$& $v_{exp}$& $M_{ej,x}/M_{\sun}$& $m$& age & $v_{core}$\\

\tableline
2.0& 1.7& 5376& 2.6& 2611& 1.65& 10& 0.26& 4900 & 1.84& 0.694& 328.7& 10086\\
\\
2.5& 2.0& 5330& 2.6& 2509& 1.67& 12& 0.28& 4924& 2.28& 0.697& 332.6& 9785\\
\\
3.0& 2.35& 5357& 2.6& 2465& 1.70& 14& 0.30& 5012& 2.71& 0.699& 331.6& 9683\\
\\
3.5& 2.67& 5353& 2.6& 2416& 1.73& 16& 0.33& 5099& 3.13& 0.699& 331.9& 9555\\
\\
4.0& 3.0& 5357& 2.6& 2384& 1.75& 18& 0.35& 5163& 3.55& 0.699& 331.8& 9474\\
\tableline
\end{tabular}
\end{center}
\tablecomments{Table columns give in order: $M_{ej}$ ejecta mass
  ($M_{\sun}$), $E_{51}$ explosion energy ($10^{51}$ ergs), $v_b$
  forward shock velocity (km s$^{-1}$), $R_b$ forward shock radius
  (pc), $v_r$ reverse shock velocity (km s$^{-1}$), $R_r$ reverse
  shock radius (pc), $\rho r_f^2$ density times forward shock radius
  at the current position of the forward shock in H atoms (or
  equivalent mass)$\times$ pc$^2$, $R_{bub}$ circumstellar bubble
  radius (pc), $v_{exp}$ ejecta expansion velocity at reverse shock
  (km s$^{-1}$), $M_{ej,x}$ mass of X-ray emitting ejecta
  ($M_{\sun}$), $m$ expansion parameter, SNR age (years), and
  $v_{core}$ expansion velocity of ejecta core-envelope boundary (km
  s$^{-1}$).  The slope of the ejecta envelope is taken to be
    n=10 from \citet{matzner99}.}
\label{tab1}
\end{table}

\begin{table}[t]
\begin{center}
\caption{Element Masses, Velocities}
\begin{tabular}{c|cc|cc|ccc}
\tableline\tableline
element& $M_{el, 2.5''}/M_{\sun}$& $M_{el, 5''}/M_{\sun}$&$v_x$& $v_y$& &
$M_{el}/M_O$& $M_{el}/M_O\vert _{solar}$\\
& & & (km s$^{-1}$)& (km s$^{-1}$) & & \\

\tableline
O & 1.80  & 2.55 & -176  &706 &&1.00& 1.00\\
Ne& 0.027 & 0.038& 1232  &486 &&0.015& 0.22\\
Mg& 0.0070&  0.0099& 247  &323 &&0.0039& 0.12\\
Si& 0.038 & 0.054 &-672  &677 && 0.021& 0.12\\
S& 0.020 & 0.028& -685  &787 && 0.011& 0.053\\
Ar& 0.010  & 0.015& -656 & 920 && 0.0056& 0.013\\
&&&&&&&\\
Fe$_{Si}$& 0.075 & 0.11 & -486  &420 && 0.041& 0.23\\
Fe$_{\alpha}$&0.023 & 0.033& -1900 &-104 && 0.016& \\
&&&&&&&\\
M$_X$& 2.00 &  2.84 & -210&  680\\
M$_{tot}$& 2.18 &  3.14 & -210& 680\\
\tableline
\end{tabular}
\end{center}
\tablecomments{Solar abundance ratios from Grevesse et
  al. (2010). Inferred masses scale as the square root of the filling
  factor.  The Fe masses are taken for a $\chi ^2$ per degree of
  freedom threshold of 1.2 as discussed in the text and given in Table
  3.}
\label{tab2}
\end{table}

\begin{table}[t]
\begin{center}
\caption{Fe$_{Si}$, Fe$_{\alpha}$ Masses}
\begin{tabular}{ccccccc}
\tableline\tableline
Fe$_{Si, 2.5''}$& Fe$_{\alpha , 2.5''}$& Fe$_{Si, 5''}$& Fe$_{\alpha , 5''}$&$\chi ^2/\nu$&
${1\over\Gamma\left(\nu /2\right)}\int _{\chi ^2/2}^{\infty}x^{\nu /2-1}\exp -xdx$& $f_{reg}$\\

\tableline
0.061& 0.092& 0.087& 0.13& & $10^0$& 1.0\\
0.072 & 0.038& 0.10& 0.054& 1.0& 0.5& 0.29\\
0.074 & 0.029& 0.10& 0.041& 1.1& $10^{-1}$& 0.21\\
0.075 & 0.023& 0.11 & 0.033& 1.2& $10^{-2}$& 0.17\\
0.075 & 0.021& 0.11 & 0.030& 1.3& $10^{-3}$& 0.15\\
0.076 & 0.018& 0.11 & 0.026& 1.37& $10^{-4}$& 0.13\\
0.076 & 0.016& 0.11 & 0022& 1.41& $10^{-5}$& 0.11\\
0.076 & 0.014& 0.11& 0.020& 1.46& $10^{-6}$& 0.10\\
0.076 & 0.013& 0.11& 0.018& 1.56& $10^{-7}$& 0.087\\
0.076 & 0.011& 0.11& 0.016& 1.60& $10^{-8}$& 0.078\\

\\
\tableline
\end{tabular}
\end{center}
\tablecomments{Fe masses in $M_{\sun}$ from incomplete Si-burning
  (Fe$_{Si}$) and from complete Si-burning and/or $\alpha$-rich freeze
  out (Fe$_{\alpha}$), for two values of the volume filling factor
  (corresponding to length scales of 2.5$''$ and 5$''$) and a number
  of threshold values for the initial single {\it vpshock} fit. Column
  5 gives this threshold value of $\chi^2/\nu$, and column 6 the
  quantity ${1\over\Gamma\left(\nu /2\right)}\int _{\chi
    ^2/2}^{\infty}x^{\nu /2-1}\exp -xdx$.  This last quantity
    gives the probability for degrees of freedom $\nu$ that the
    initial fit used the correct model, but only with strict
    requirements that are not met here (see the text). The last
  column gives the fraction $f_{reg}$ of the 2982 regions considered
  that were fitted with the composite {\it vpshock + NEI} model used
  in the Fe mass calculation. }
\label{tab3}
\end{table}

\begin{figure}
\label{fig:images}
\centerline{\includegraphics[scale=0.45]{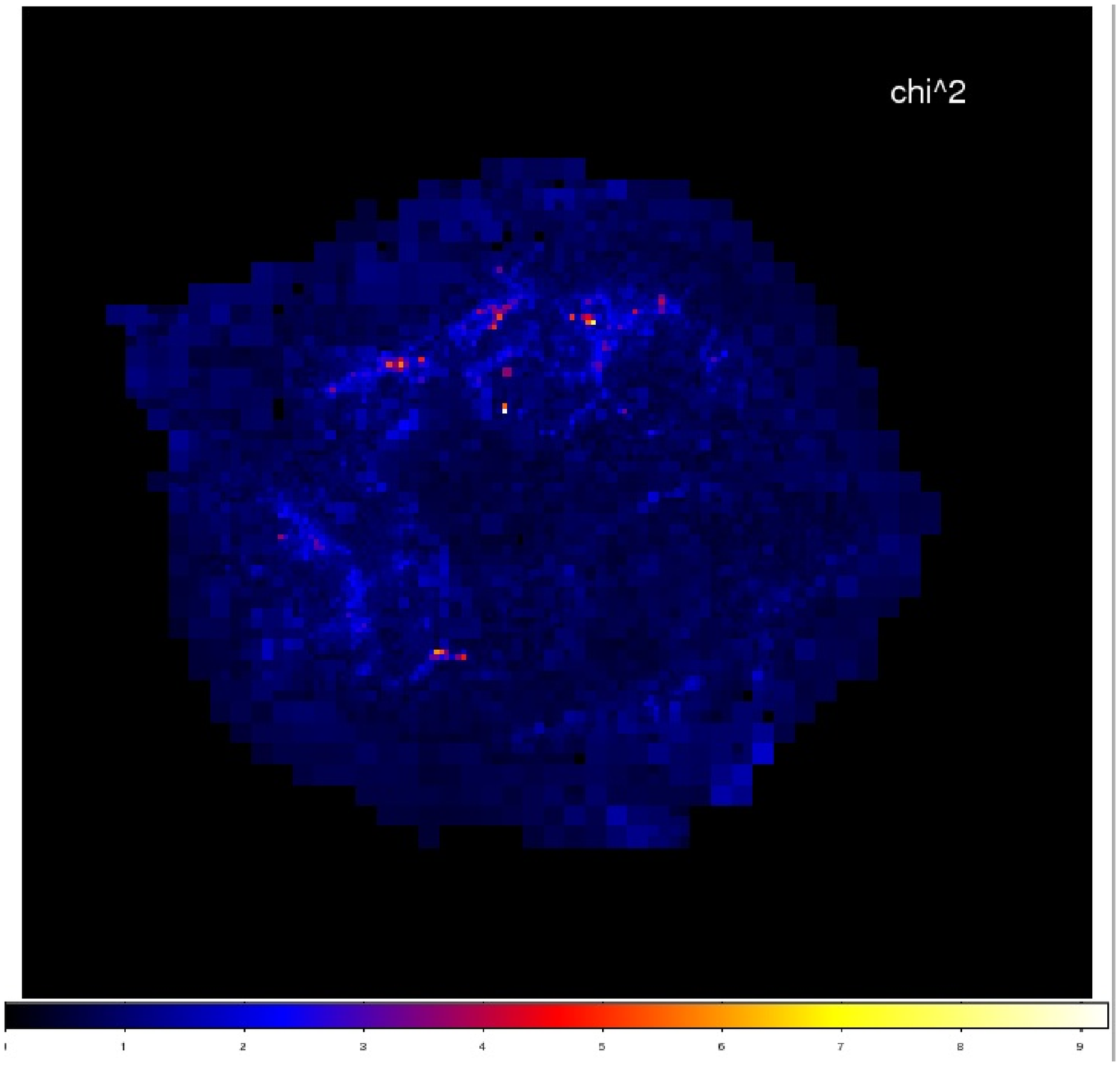}\includegraphics[scale=0.45]{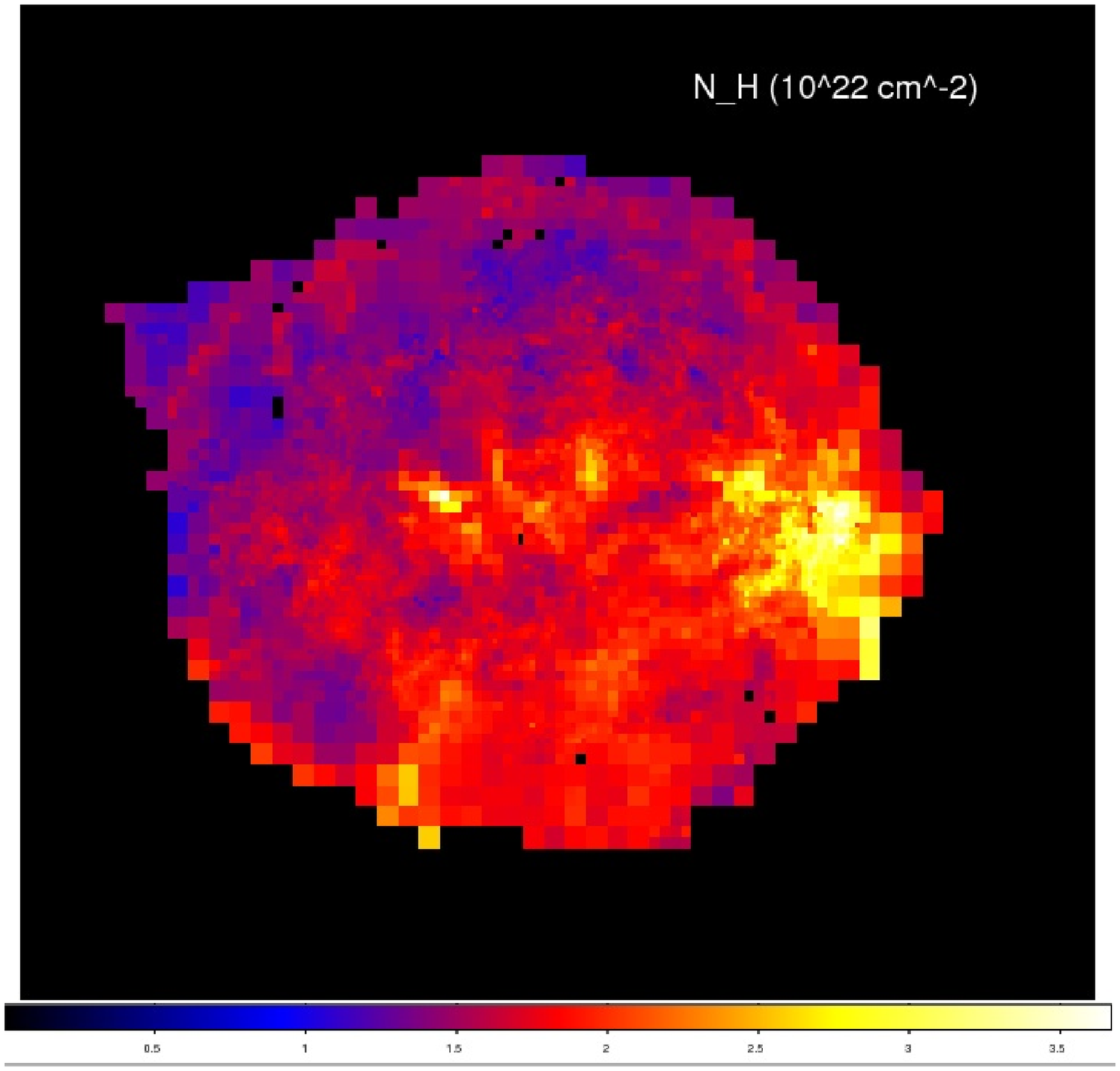}}
\centerline{\includegraphics[scale=0.45]{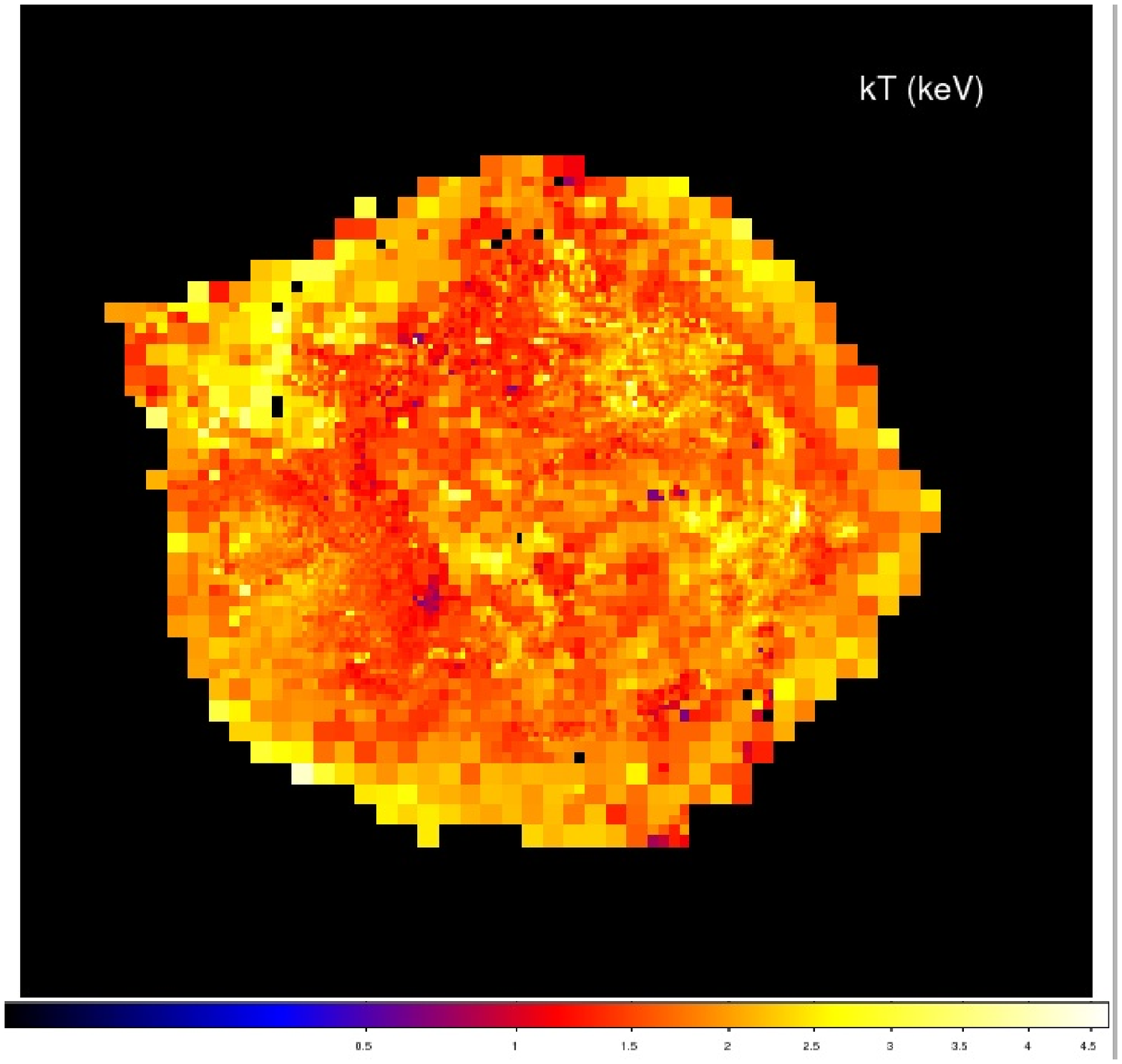}\includegraphics[scale=0.45]{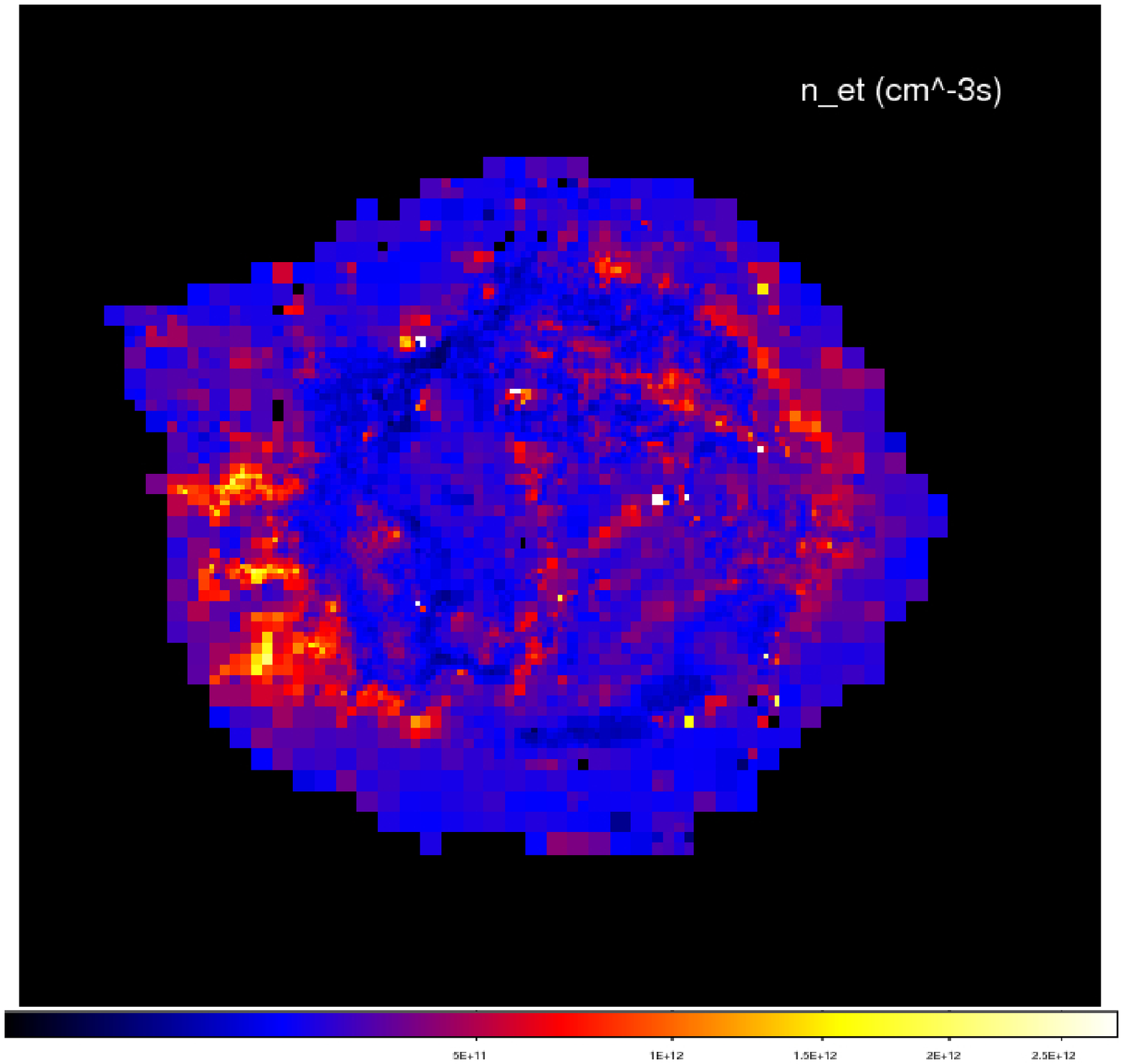}}
\centerline{\includegraphics[scale=0.45]{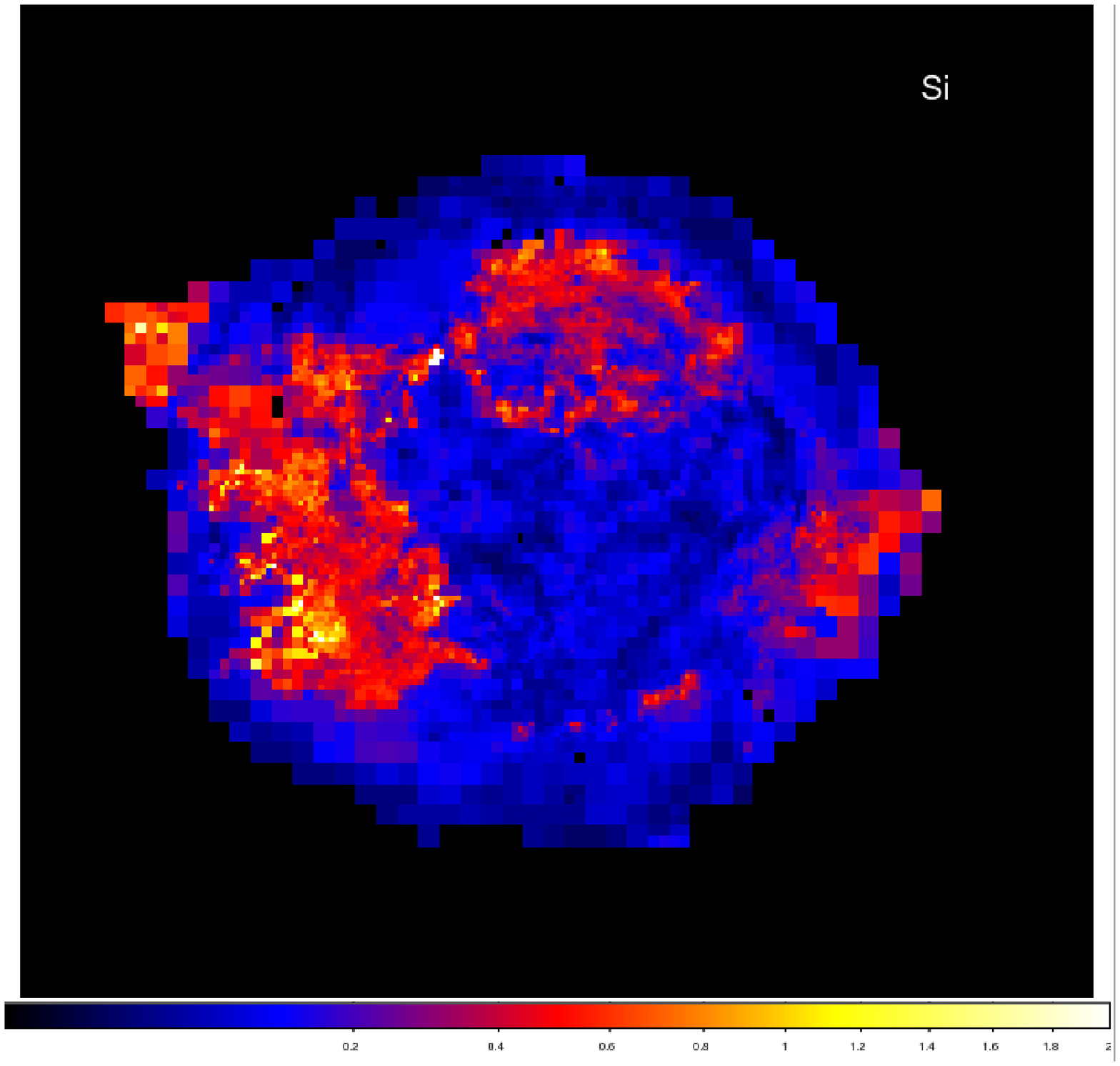}\includegraphics[scale=0.45]{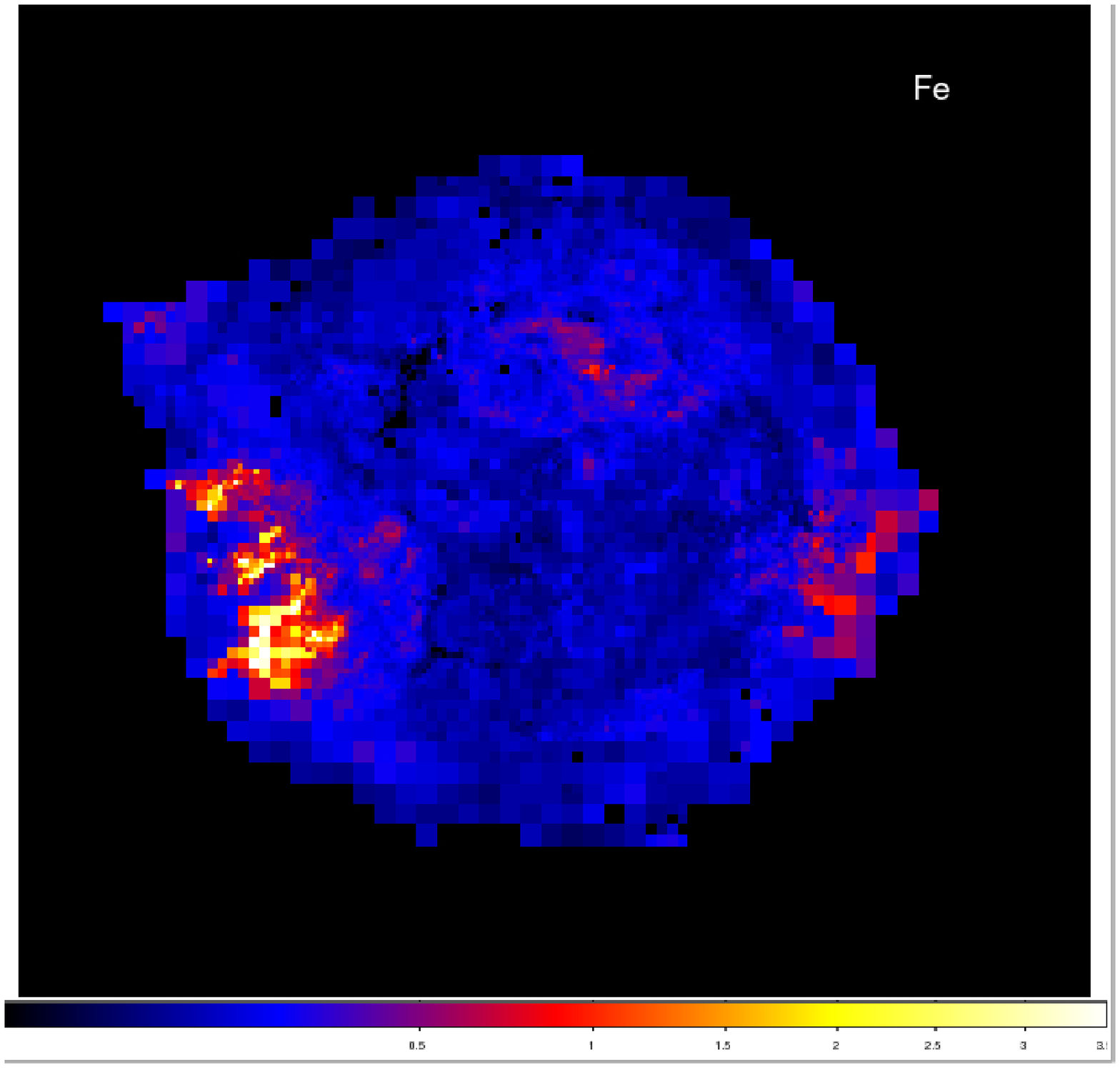}}
\end{figure}
\begin{figure}
\setcounter{figure}{0}
\centerline{\includegraphics[scale=0.45]{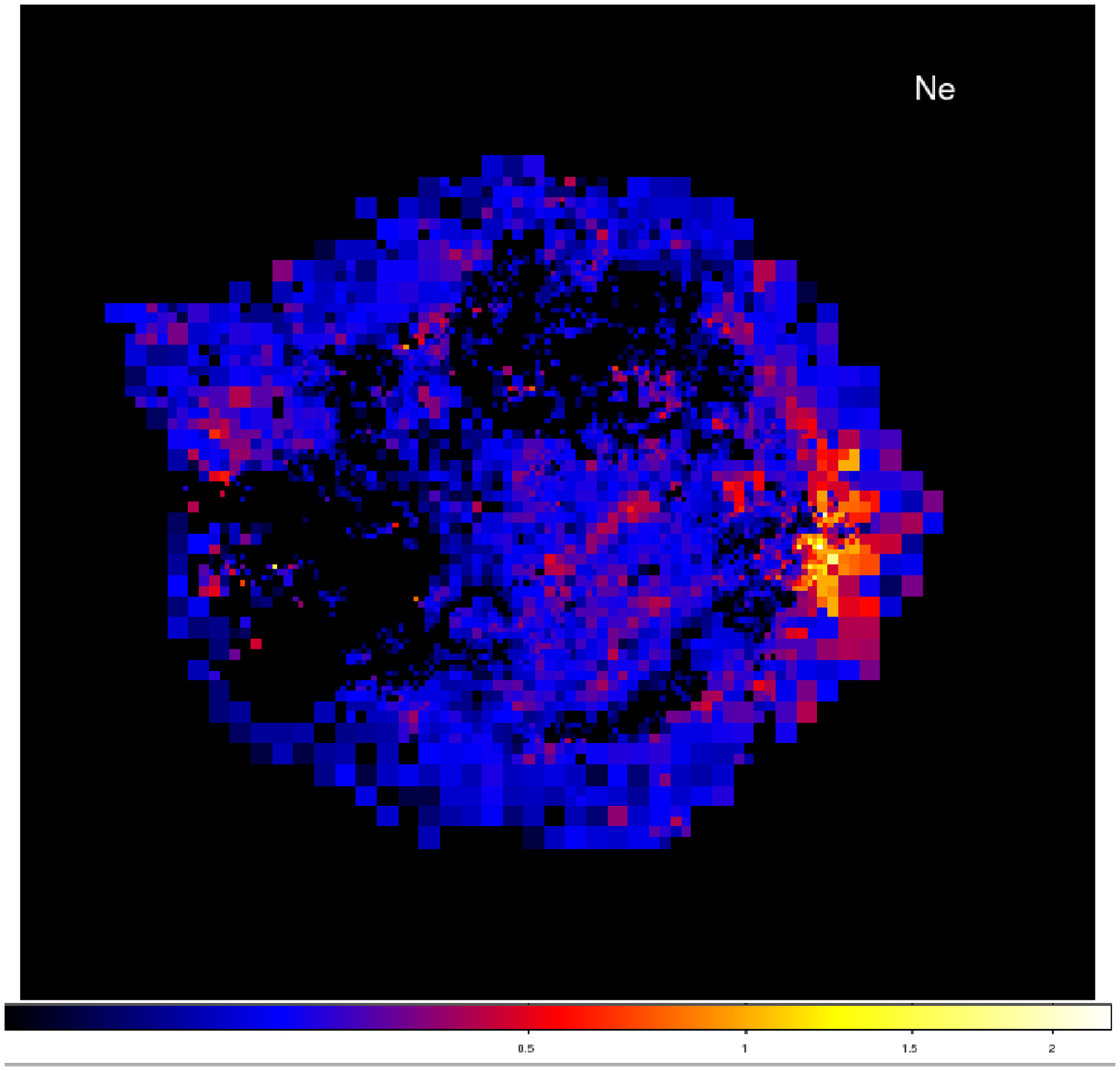}\includegraphics[scale=0.45]{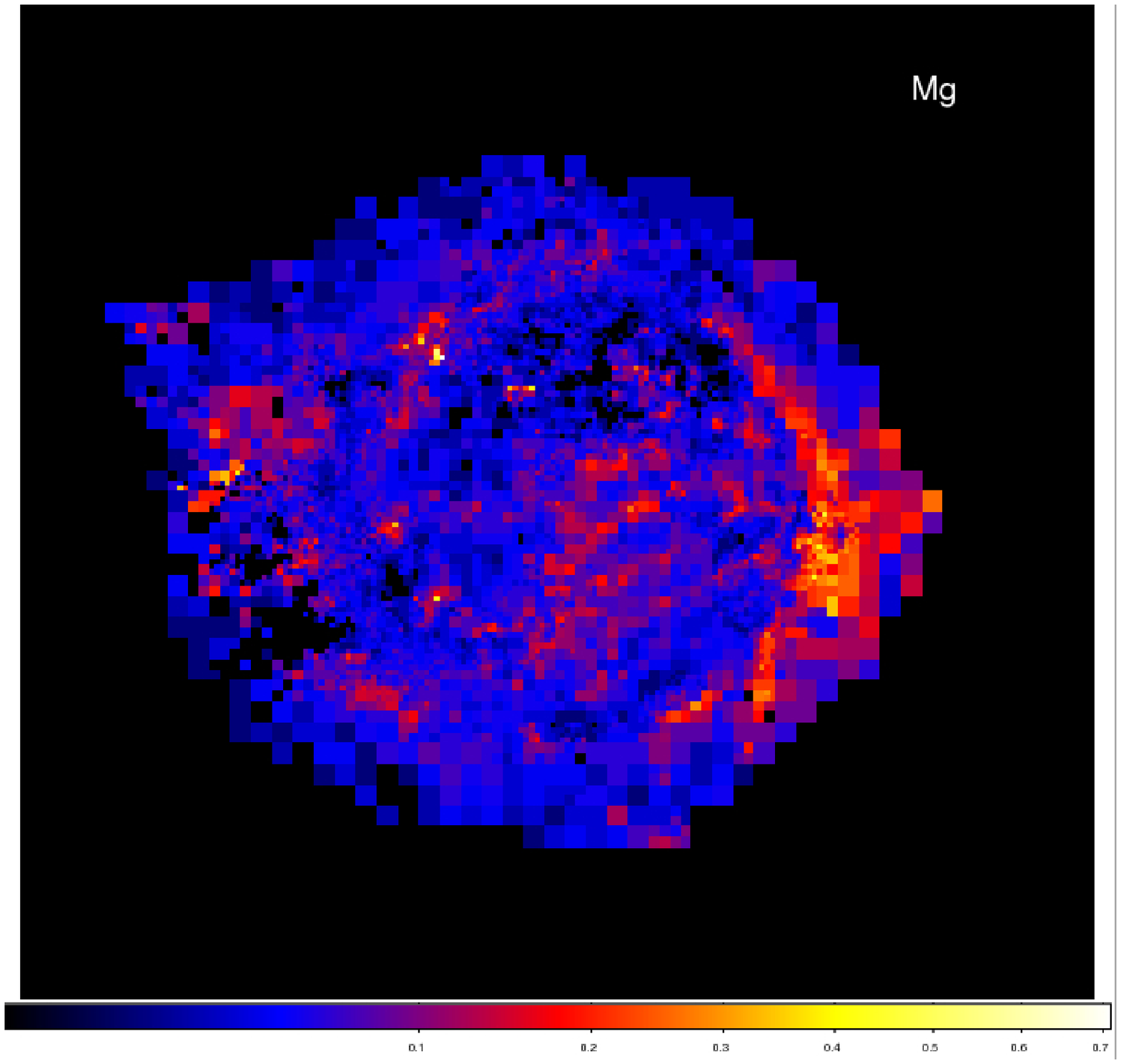}}
\centerline{\includegraphics[scale=0.45]{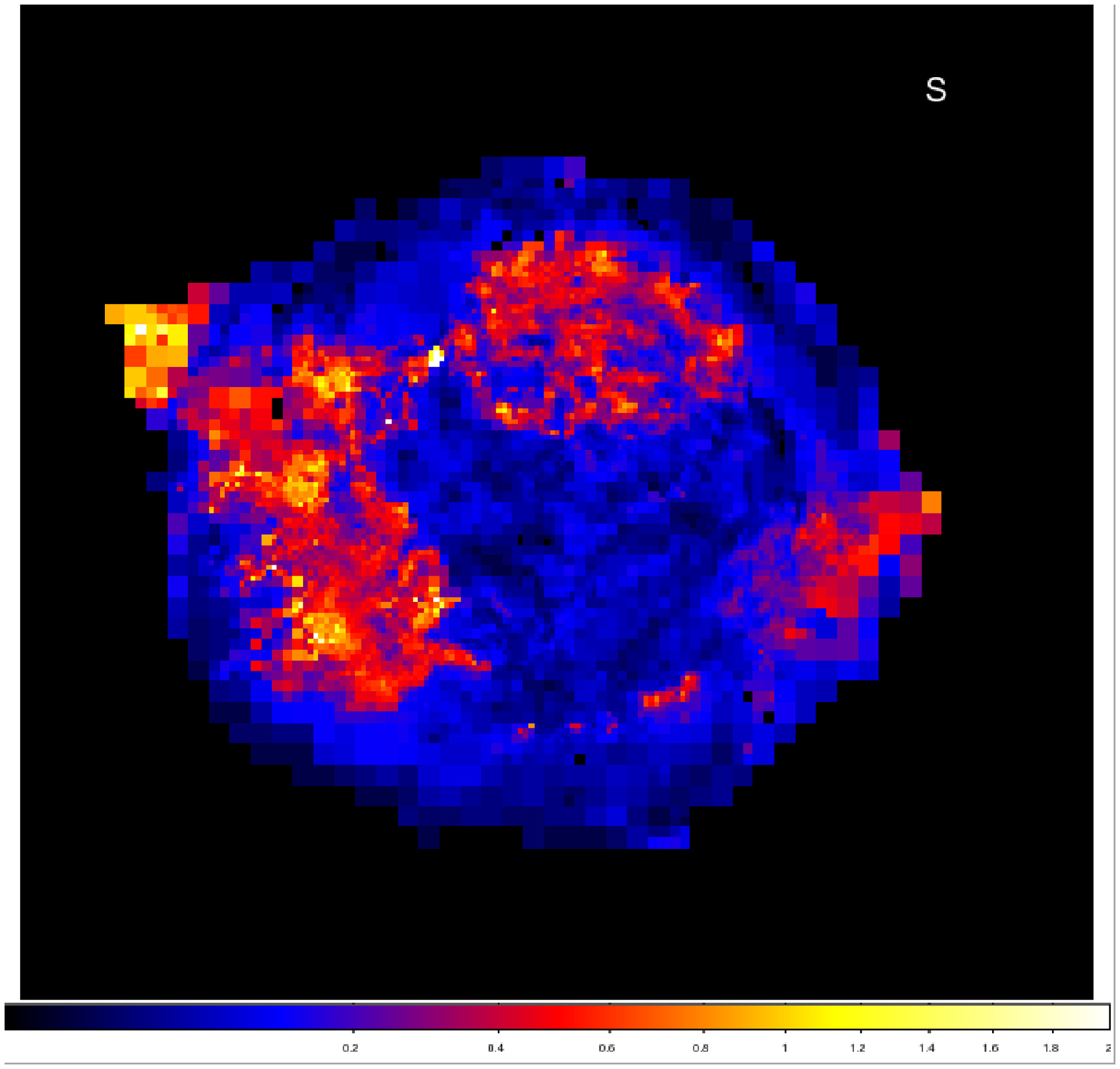}\includegraphics[scale=0.45]{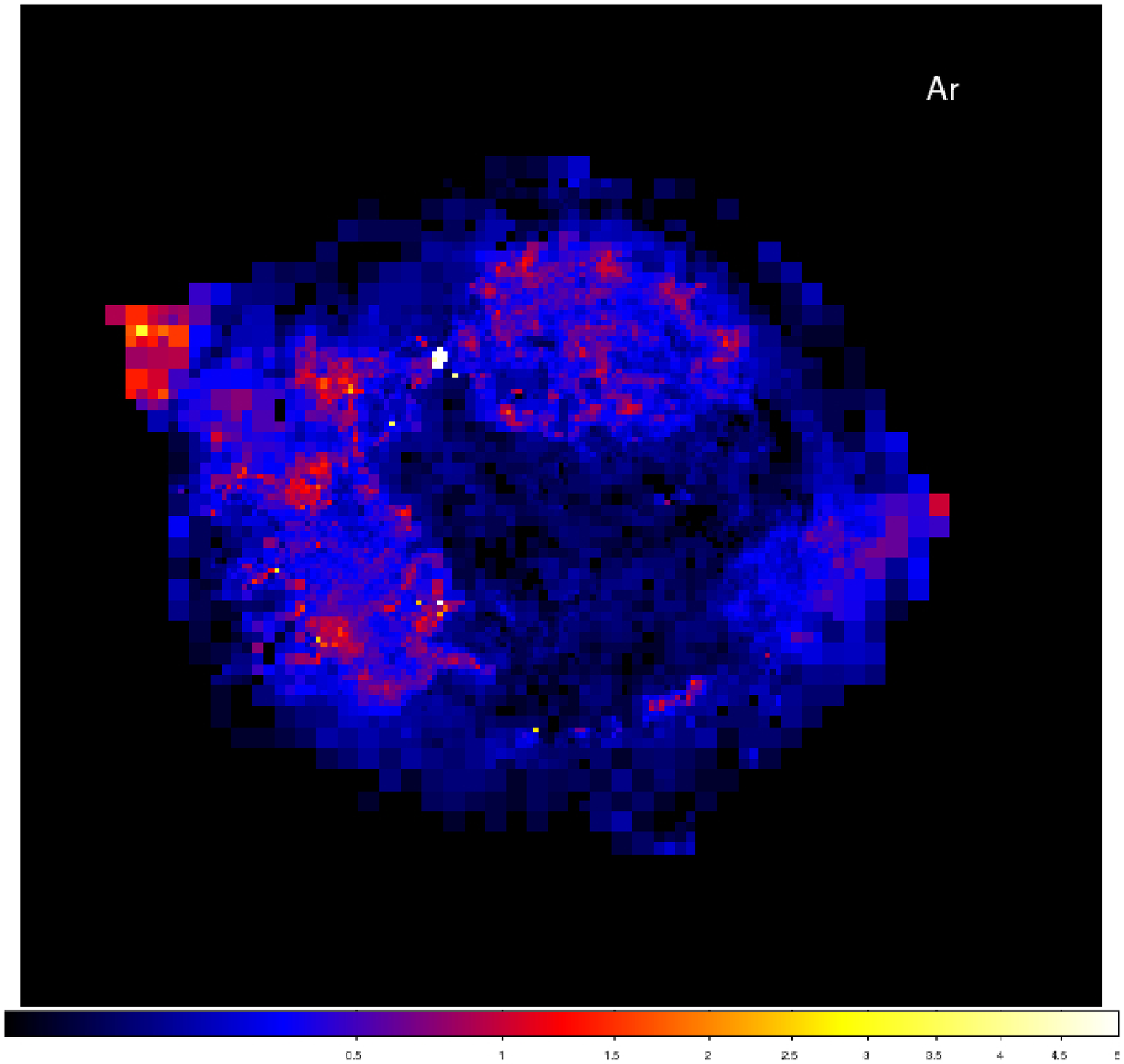}}
\figcaption{Partial set of fitted parameters for the single {\em
    vpshock} model fits.  Top row: Maps of $\chi^2$ per degree of
  freedom (left), and of fitted column density N$_{\rm H}$ in
  10$^{21}$ cm$^2$ (right). Second row: Maps of temperature kT in keV
  and ionization age n$_{\rm e}$t in cm$^{-3}$s (sqrt scale). Third
  row: Maps of fitted Si and Fe element abundances, relative to the
  solar values of Anders \& Grevesse (1989) by number, both on sqrt
  intensity scale, truncated at the high end.  Bottom rows: Fitted
  abundances of the remaining elements, Ne, Mg, S, Ar.  The
  intensities are square-root scale and truncated at the high end for
  S and Ar.}


%
\end{figure}

\begin{figure}
\label{fig:vpc}
\centerline{\includegraphics[scale=0.40]{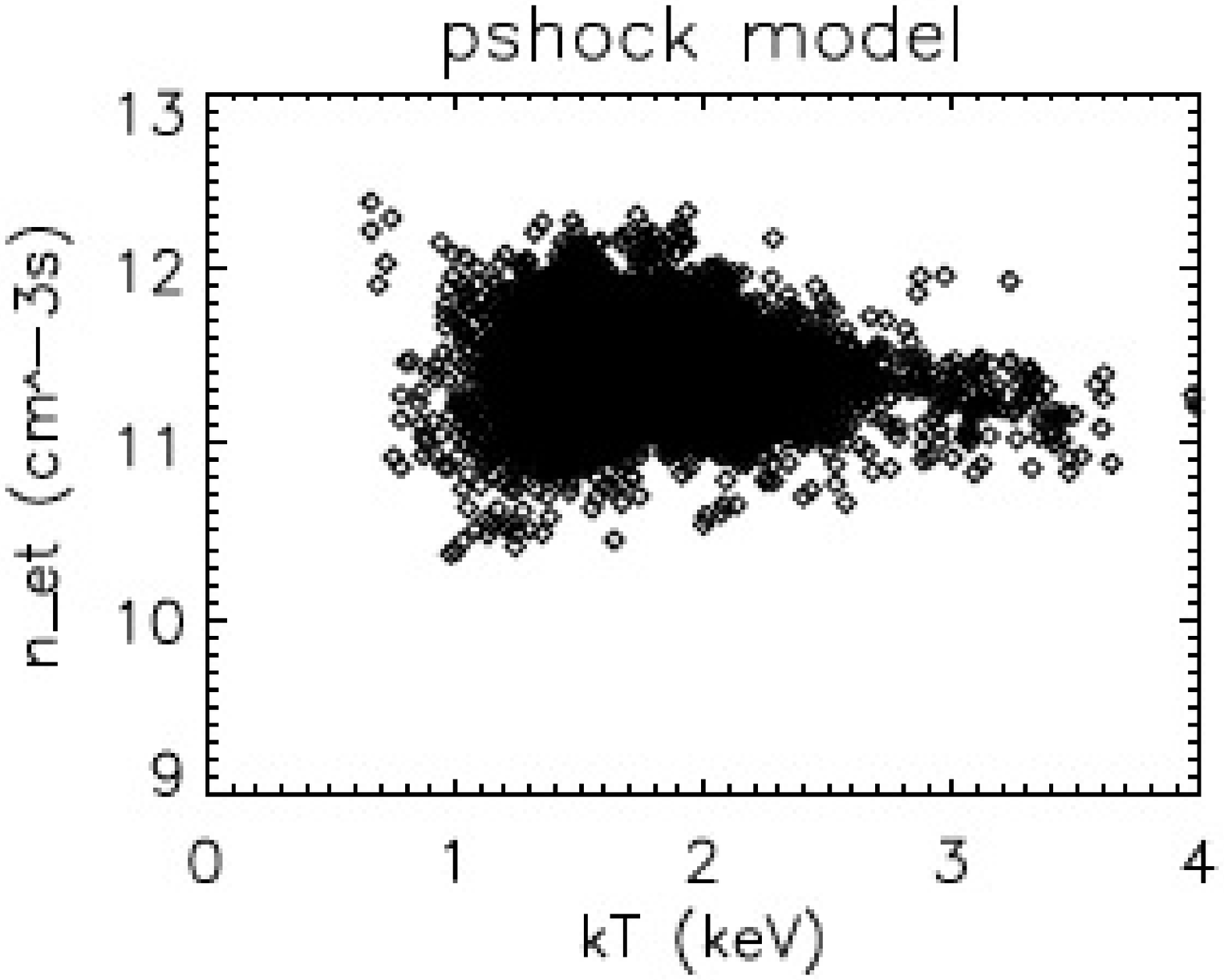}}
\centerline{\includegraphics[scale=0.40]{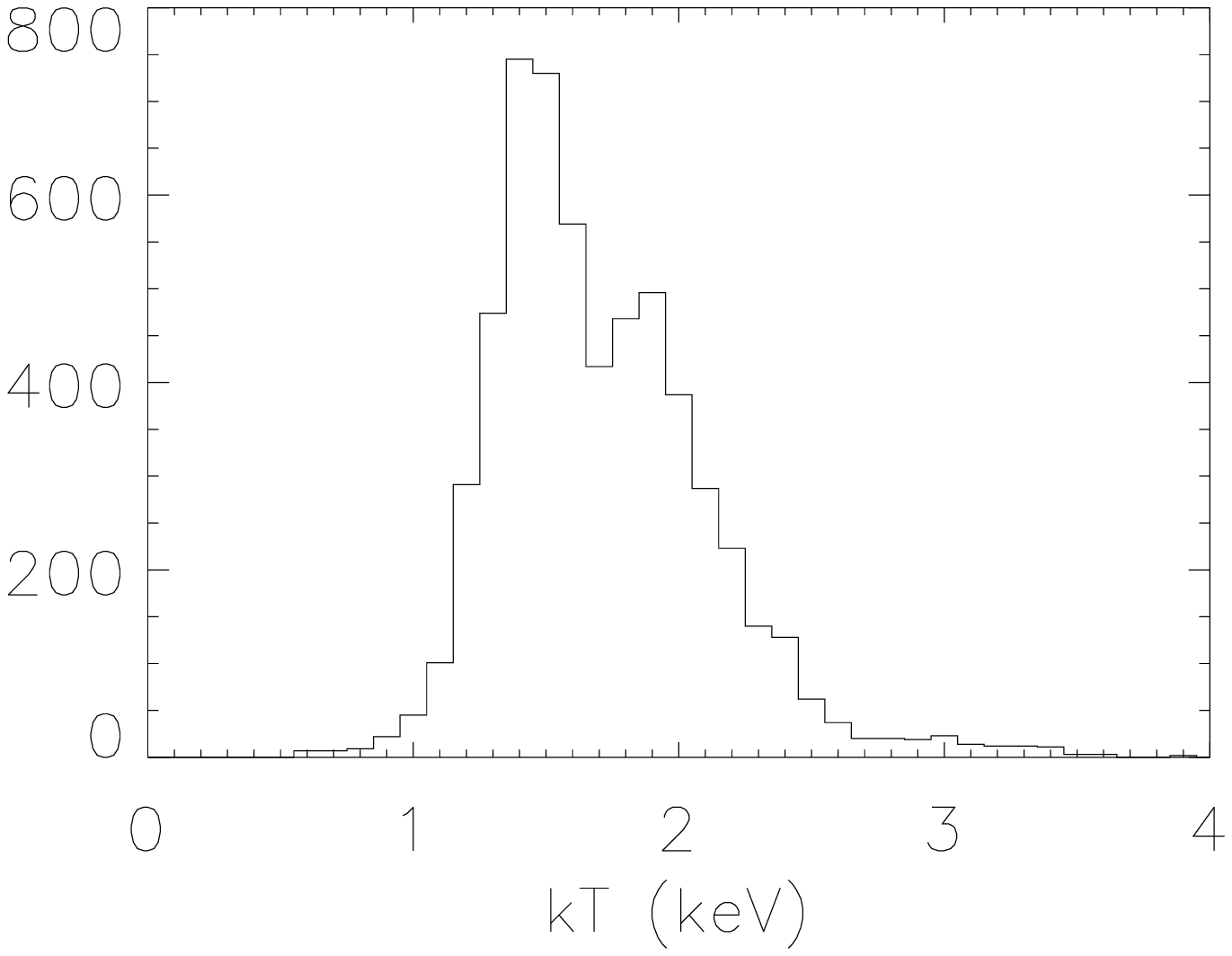}\includegraphics[scale=0.40]{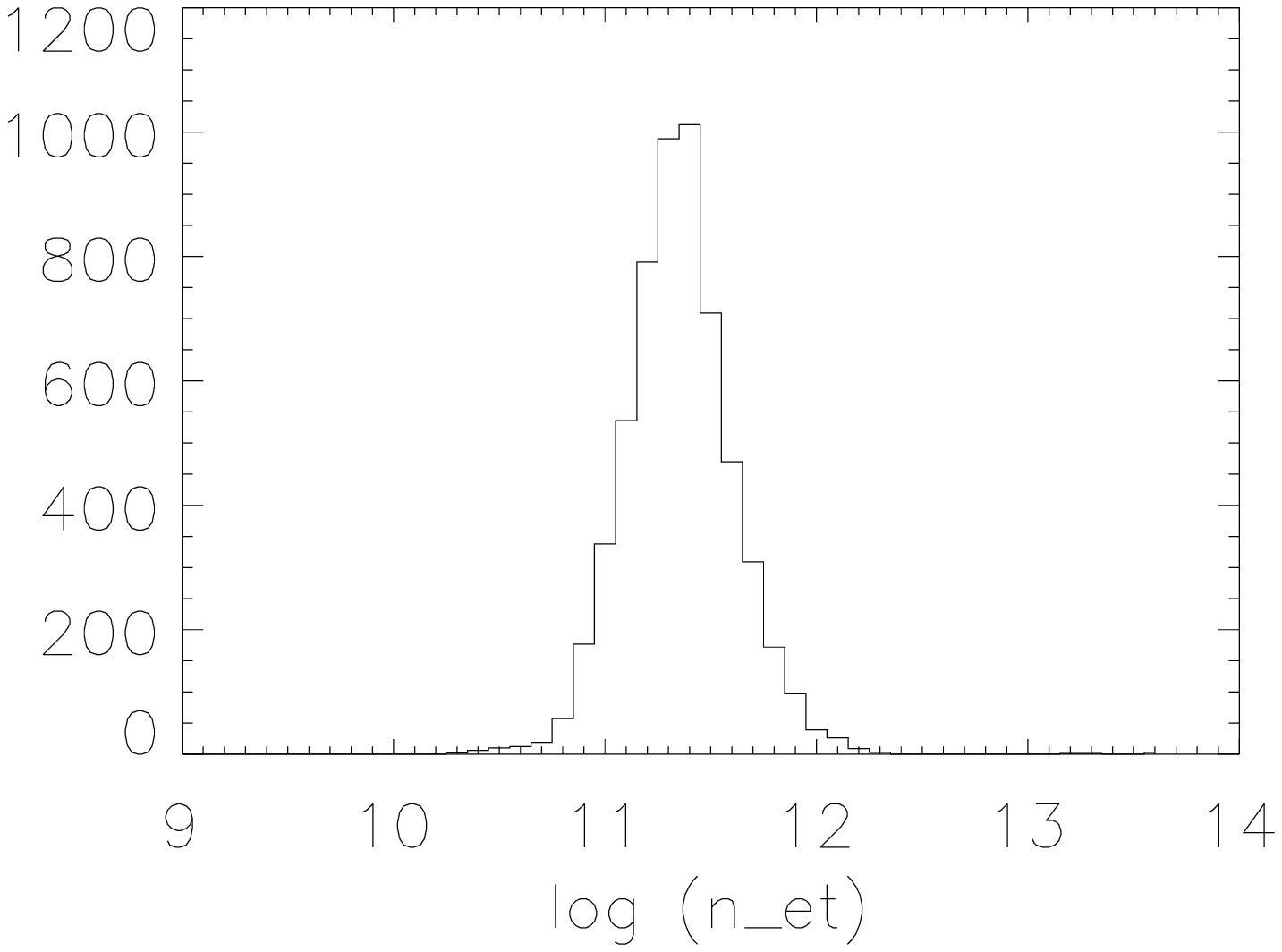}}
\figcaption{The distribution of temperature $kT$ and ionization age
  $n_et$ amongst the 6202 spectral regions fitted with {\em pshock}
  models (see Figure 1) : (top) two-dimensional distribution;
  (lower left) histogram for temperature {\em kT}; (lower right) histogram for
  ionization age {\em $n_et$}}
\end{figure}

\begin{figure}
\label{fig:fs}
\centerline{\includegraphics[scale=0.45]{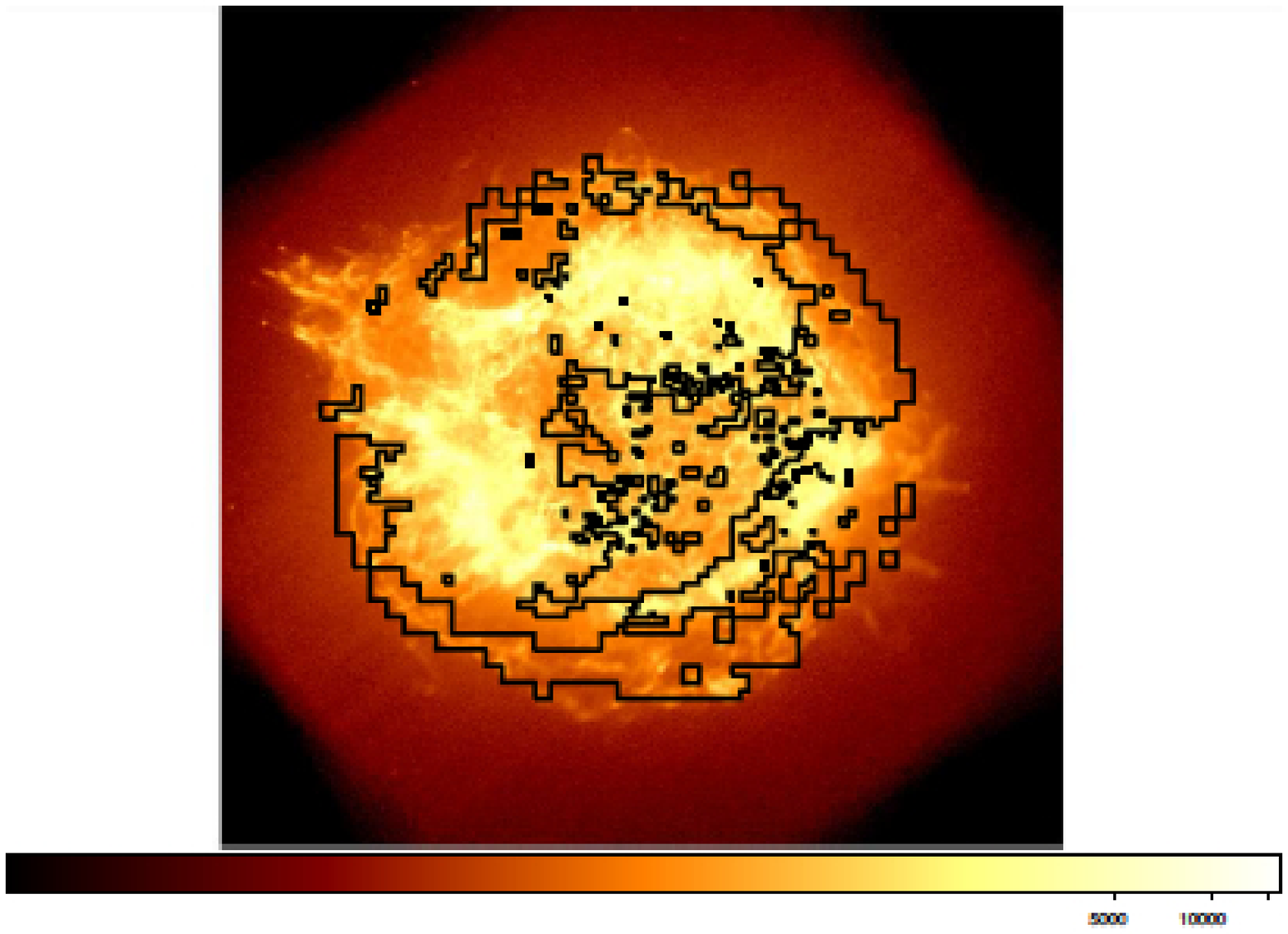}}
\centerline{\includegraphics[scale=0.40]{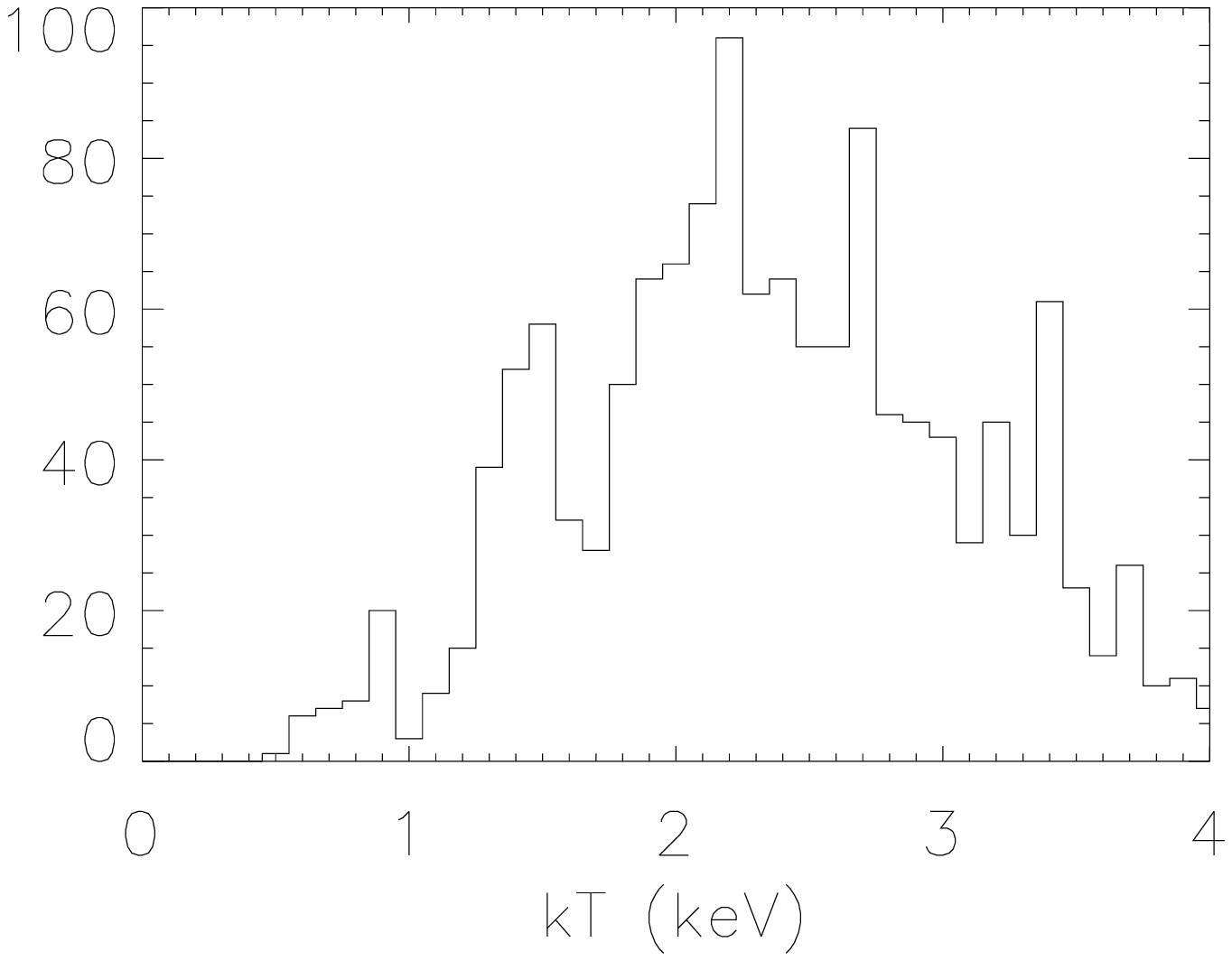}\includegraphics[scale=0.40]{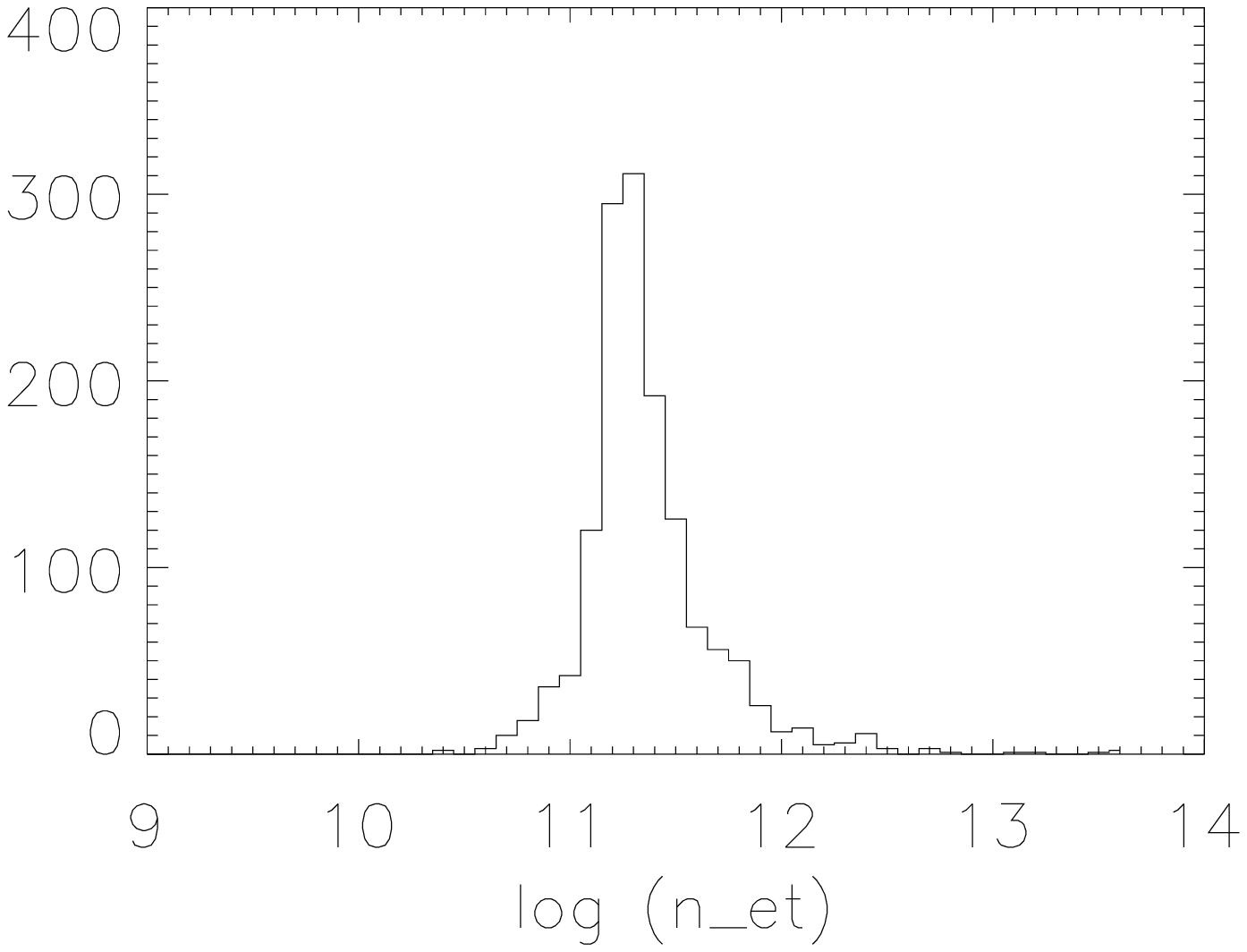}}
\caption{(top) Contours showing the 1415 regions that have been
  associated with the forward shock, and hence excluded for the ejecta
  mass calculation, based on the merged QSF and QSF+PL fits.  For
  these forward shock regions is shown the distribution of fitted
  (lower left) temperature $kT$ (the scale is truncated above 10 keV), and
  (lower right) log$_{10}$ ionization age $n_et$.}
\end{figure}

\begin{figure}
\centerline{\includegraphics[scale=0.40]{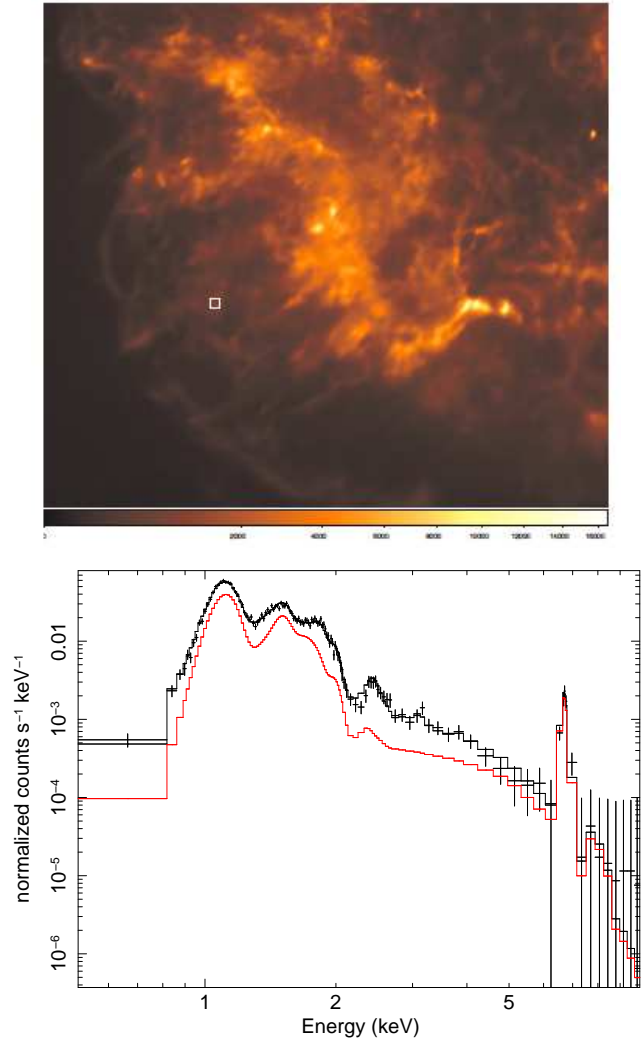}}
\centerline{\includegraphics[scale=0.35,angle=-90]{f4b.ps}}
\caption{(Top) Southeast region of Cas A showing the extraction region used here
  for the pure Fe cloud of \citet{hwang03}.  (Bottom) Fitted spectrum,
  in which the local background has been modelled rather than
  subtracted and the unbinned spectrum fitted using C-statistics.  The
  spectrum is shown binned for clarity, with the black trace giving
  the total source plus background spectrum, and the red trace giving
  only the source spectrum.}
\end{figure}

\begin{figure}
\label{fig:spectra}
\centerline{\includegraphics[scale=0.35,angle=-90]{f5a.ps}\includegraphics[scale=0.35,angle=-90]{f5b.ps}}
\centerline{\includegraphics[scale=0.35,angle=-90]{f5c.ps}}
\centerline{\includegraphics[scale=0.35,angle=-90]{f5d.ps}\includegraphics[scale=0.35,angle=-90]{f5e.ps}}
\end{figure}
\begin{figure}
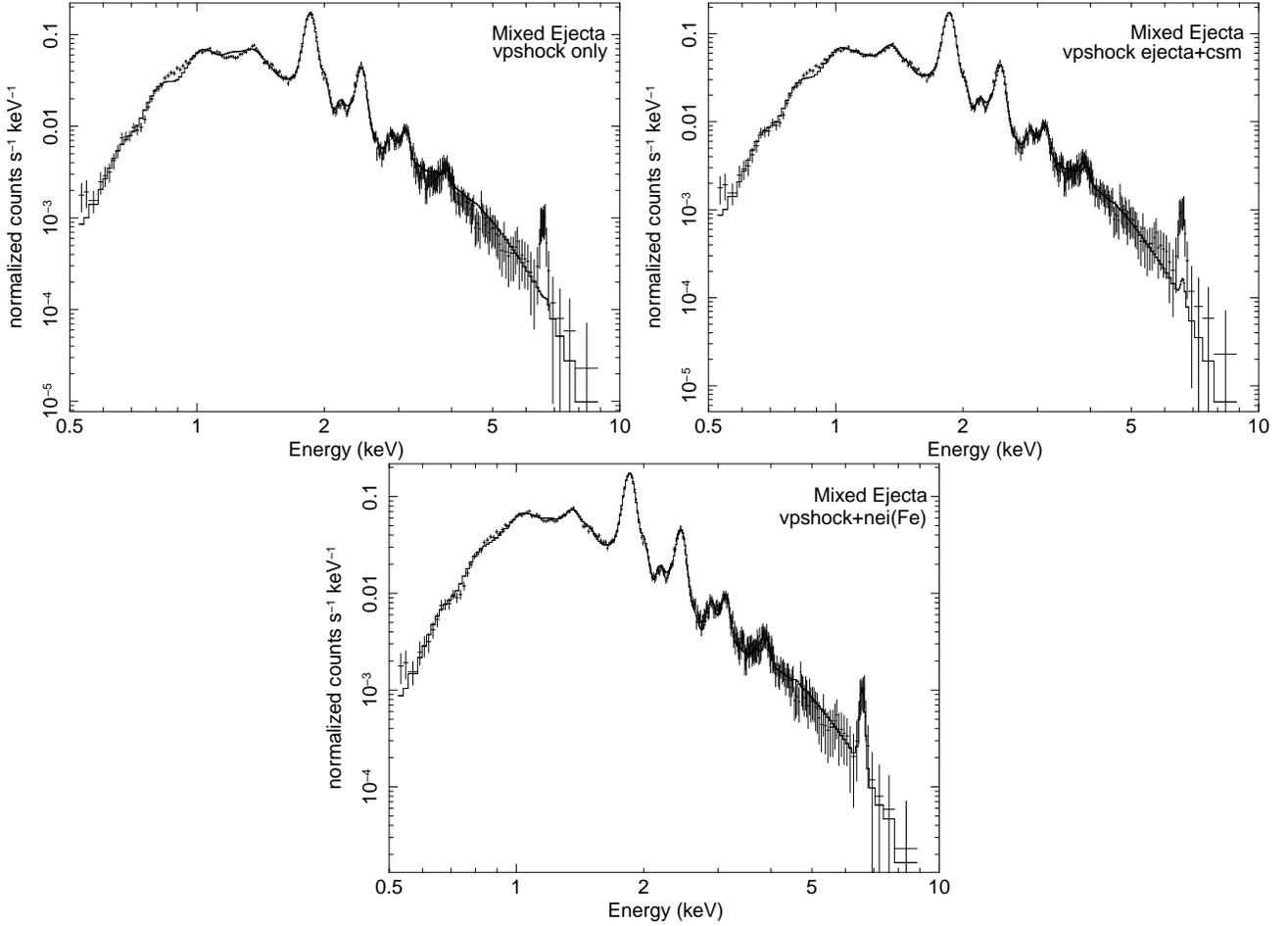

\setcounter{figure}{4}
\centerline{\includegraphics[scale=0.35,angle=-90]{f5f.ps}\includegraphics[scale=0.35,angle=-90]{f5g.ps}}
\centerline{\includegraphics[scale=0.35,angle=-90]{f5h.ps}}
\caption{Sample spectra exemplifying typical spectral types seen in
  Cas A (first two rows) Spectra associated with the forward shock,
  showing various mixtures of thermal emission associated with shocked
  circumstellar medium and nonthermal emission.  (third row) Typical
  ``Si-dominated'' and Fe-dominated ejecta spectra.  (last two rows)
  For the same spectrum, a comparison of single-component {\it
    vpshock} ejecta models, ejecta plus shocked CSM models, and a
  model with two ejecta components corresponding to ``normal'' ejecta
  and ``pure Fe'' ejecta.  For reference, this spectrum had a
    $\chi^2/\nu = 1.98$ for the single component {\em vpshock} model
    fit.}
\end{figure}

\begin{figure}
\centerline{\includegraphics[scale=0.35]{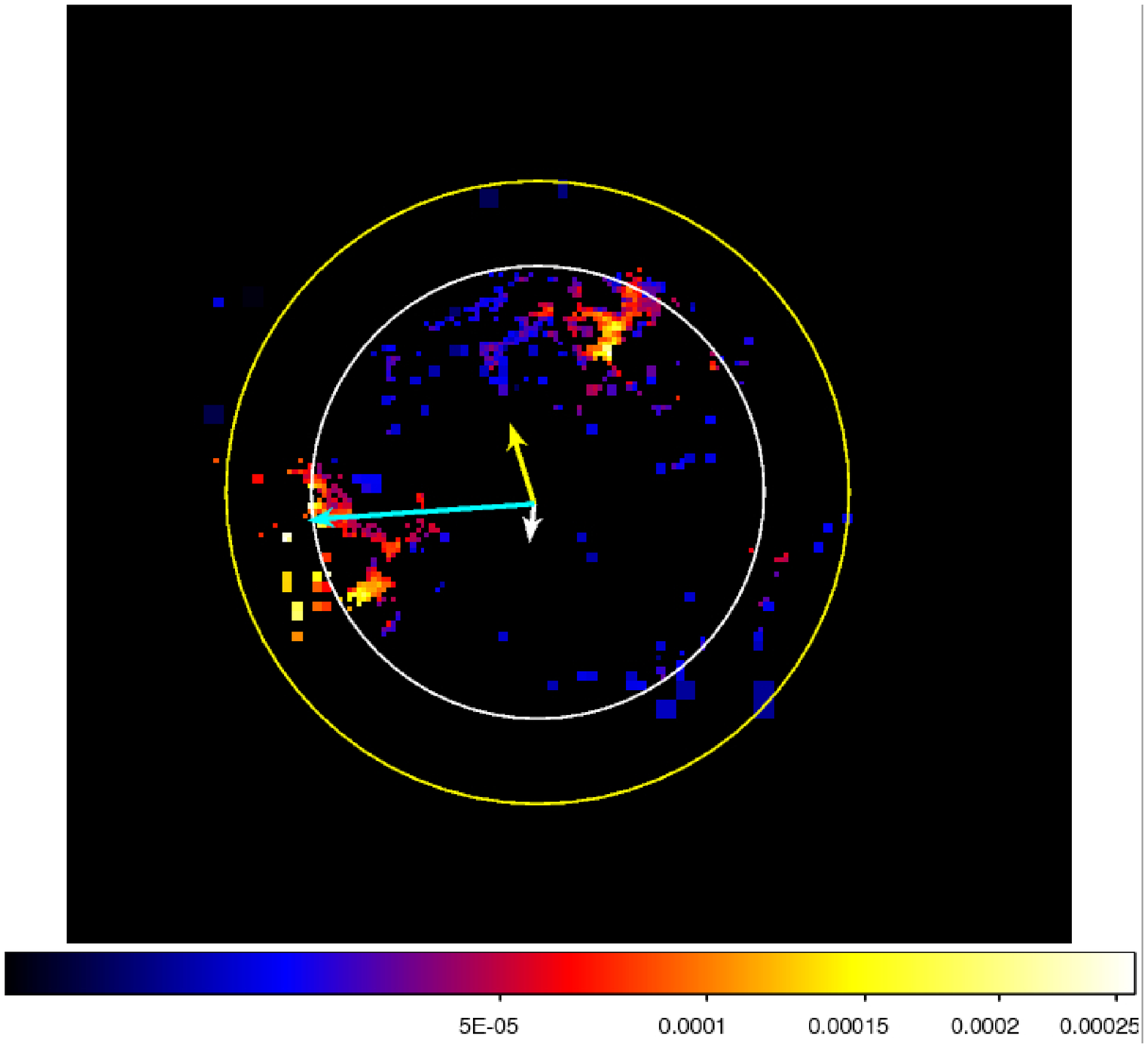}
\includegraphics[scale=0.35]{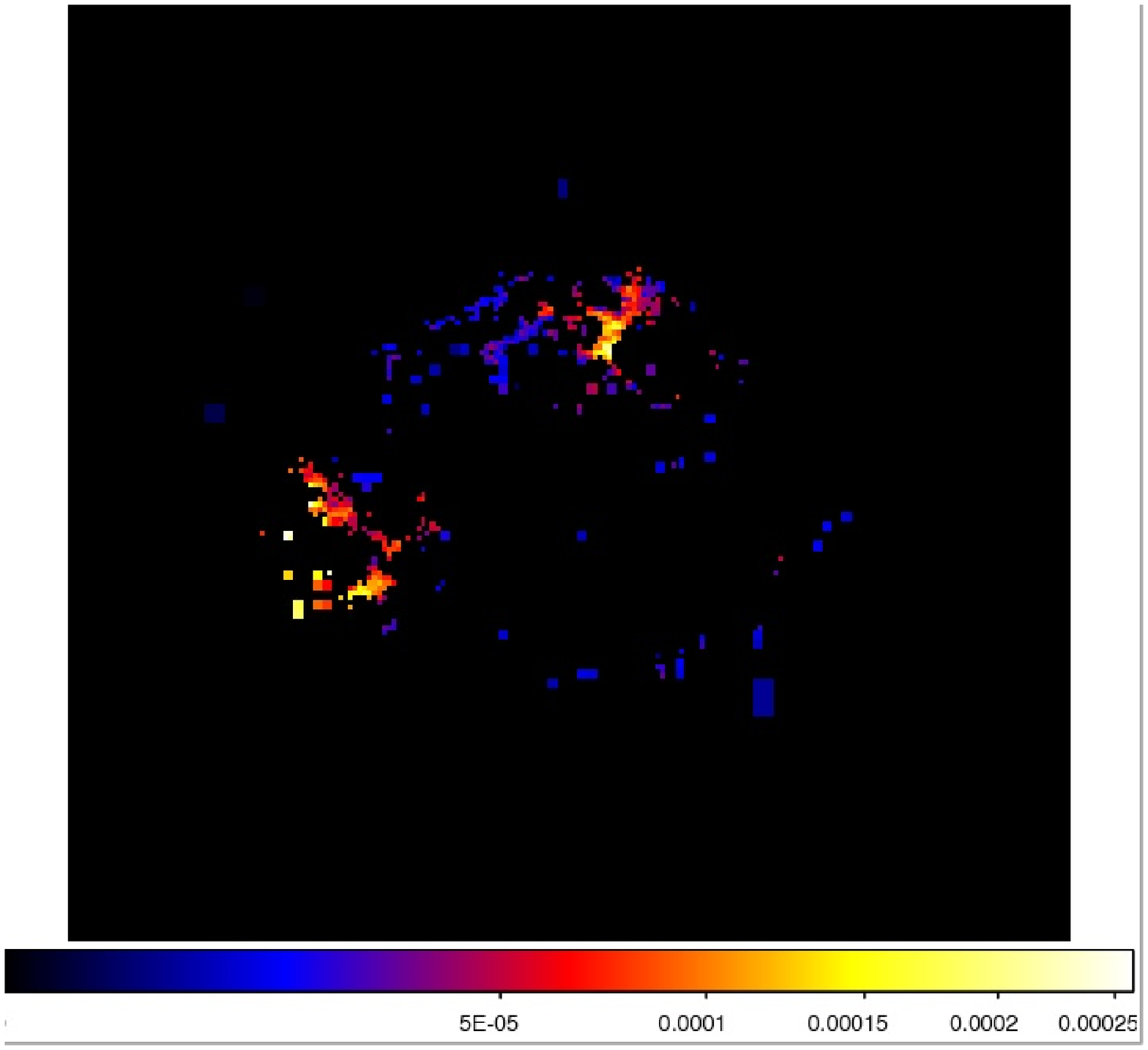}
\includegraphics[scale=0.35]{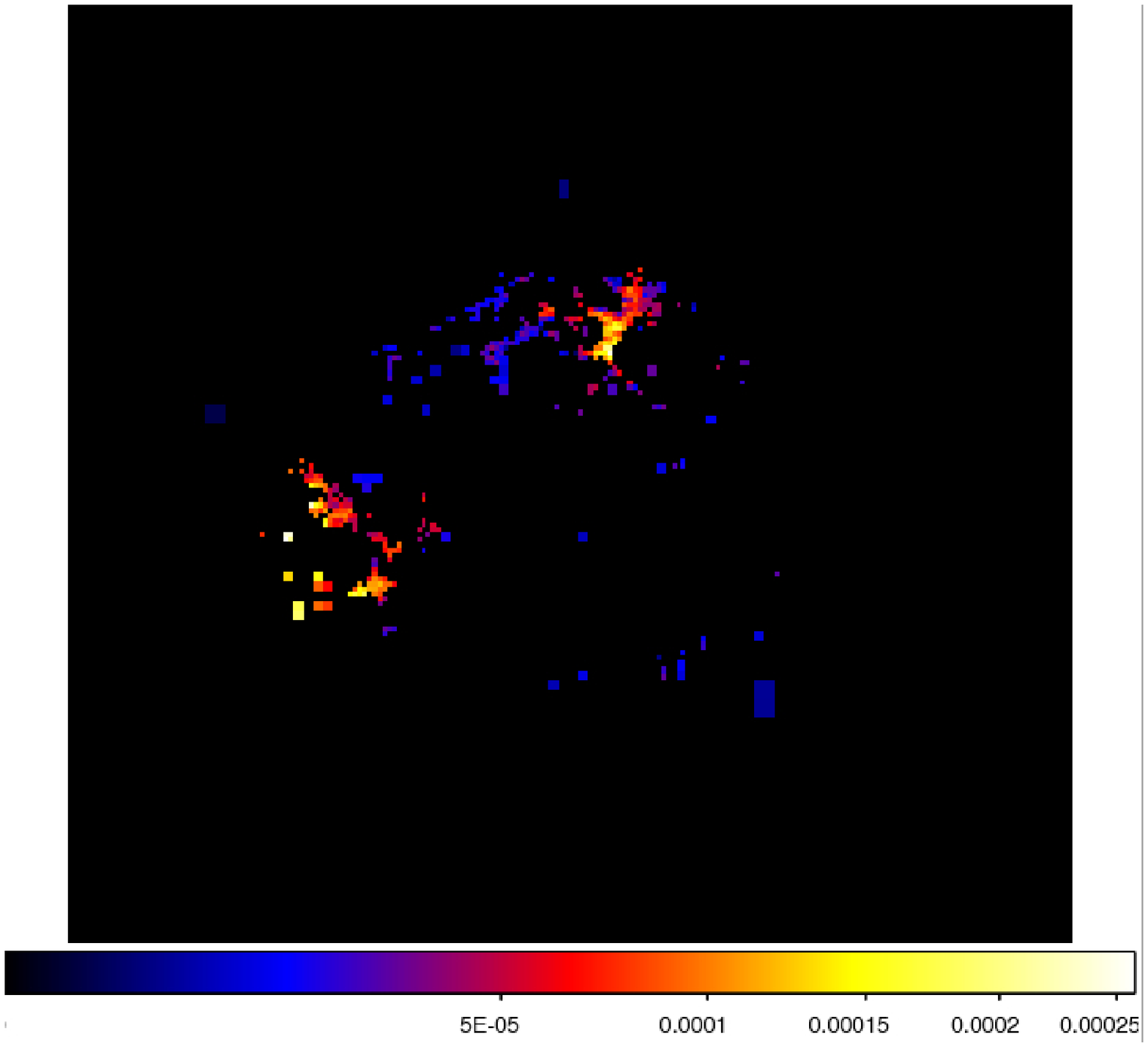}}
\caption{The distribution of the ``pure'' Fe ejecta component in
  two-component {\em vpshock + NEI} model fits where the
    $\chi^2/\nu$ cutoff thresholds for a single-component {\em
    vpshock} model fit were 1.1, 1.2, and 1.3 (left to right).  In
  the left panel, we show fiducial circles showing the radius and
  position of the forward shock (2.6 pc at 3.4 kpc distance, in
  yellow) and the contact discontinuity (1.9 pc, in white), and the
  velocity vectors for the NS (white), the ``pure'' Fe ejecta (light
  blue), and total ejecta (yellow) from Table 2 (see subsection 4.2).
  The circles and velocity vectors are centered at the explosion
  center of \citet{thorstensen01}.}
\end{figure}

\end{document}